\documentclass[twocolumn]{aastex62}
\usepackage{natbib}
\usepackage{comment}
\usepackage{amsmath}	
\usepackage{amssymb}
\usepackage{footnote}
\usepackage{appendix}
\usepackage[graphicx]{realboxes}
\usepackage{threeparttable}

\newcommand{\ha}{H$\alpha$}

\graphicspath{{./}{figures/}}
\def\ltp{\left ( \,}
\def\rtp{\, \right  ) }

\newcommand{\mkms}{{\rm km \, s^{-1}}}

\newcommand{\rafrb}{22h16m04.77s}   
\newcommand{\decfrb}{-07d53m53.7s}  
\newcommand{\stataerr}{0.19}  
\newcommand{\sysaerr}{0.18}  
\newcommand{\statberr}{0.18}  
\newcommand{\sysberr}{0.18}  

\newcommand{\hgname}{HG\,190608}
\newcommand{\sdssname}{SDSS\,J221604.90-075356.0}
\newcommand{\zhg}{0.11778}  

\newcommand{\mmuhbeta}{\mu_{\rm H\beta}}  
\newcommand{\muhbeta}{$\muhbeta$}
\newcommand{\mmuchalpha}{\mu^{\rm corr}_{\rm H\alpha}}  
\newcommand{\muchalpha}{$\mmuchalpha$}
\newcommand{\sbhbeta}{\mathrm{(3.36\pm0.21)\times10^{-17}\;erg\;s^{-1}\;cm^{-2}\;arcsec^{-2}}}  
\newcommand{\sbchalpha}{\mathrm{(38.6\pm2.4)\times10^{-17}\;erg\;s^{-1}\;cm^{-2}\;arcsec^{-2}}}  
\newcommand{\memhalpha}{{\rm EM}_{\rm H\alpha}}  
\newcommand{\emhalpha}{$\memhalpha$}

\newcommand{\iangle}{37 \pm 3^\circ}
\newcommand{\vcirc}{141 \pm 8\;\mkms}  
\newcommand{\dv}{-79 \pm 0.52\;\mkms} 
\newcommand{\vsigma}{15 \pm 0.55\;\mkms}  

\newcommand{\mmstar}{M_\star}

\newcommand{\mmsun}{{M}_\odot}
\newcommand{\msun}{$\mmsun$}
\newcommand{\mmhalo}{M_{\rm halo}}
\newcommand{\mhalo}{$\mmhalo$}

\newcommand{\mdmunits}{{\rm pc \, cm^{-3}}} 
\newcommand{\dmunits}{$\mdmunits$}
\newcommand{\dmval}{339.79}  

\newcommand{\mdmfrb}{{\rm DM}_{\rm FRB}}
\newcommand{\dmfrb}{$\mdmfrb$}
\newcommand{\mdmmwism}{{\rm DM}_{\rm MW,ISM}}

\newcommand{\mdmhost}{{\rm DM}_{\rm Host}}
\newcommand{\dmhost}{$\mdmhost$}
\newcommand{\mdmismhost}{{\rm DM}_{\rm Host,ISM}}
\newcommand{\dmismhost}{$\mdmismhost$}
\newcommand{\mdmhalohost}{{\rm DM}_{\rm Host,Halo}}
\newcommand{\dmhalohost}{$\mdmhalohost$}

\newcommand{\msfrsurf}{\Sigma_{\rm SFR}}
\newcommand{\sfrsurf}{$\msfrsurf$}

\newcommand{\rmvalue}{353}  
\newcommand{\rmerr}{2}  
\newcommand{\rmunits}{{\rm rad \, m^{-2}}}

\newcommand{\mdt}{\tau}
\newcommand{\dt}{$\mdt$}
\newcommand{\dtval}{3.3} 
\newcommand{\dterr}{0.2}
\newcommand{\dtunits}{ms}

\newcommand{\hst}{{\it HST}} 

\received{April 27, 2020}
\revised{September 11, 2021}
\accepted{September 16, 2021}

\shorttitle{FRB~190608 Host Galaxy}
\shortauthors{Chittidi et al.}

\begin{document}

\title{Dissecting the Local Environment of FRB 190608 in the Spiral Arm of its Host Galaxy
}

\correspondingauthor{Jay S. Chittidi}
\email{jay.chittidi@colorado.edu
}
\author{Jay S. Chittidi}
\affil{Maria Mitchell Observatory, 4 Vestal Street, Nantucket, MA 02554, USA}
\affil{Department of Astrophysics and Planetary Sciences, University of Colorado, Boulder, CO 80309, USA}

\author{Sunil Simha}
\affil{University of California - Santa Cruz, 1156 High St., Santa Cruz, CA, USA 95064}

\author{Alexandra Mannings}
\affil{University of California - Santa Cruz, 1156 High St., Santa Cruz, CA, USA 95064}

\author{J. Xavier Prochaska}
\affil{University of California - Santa Cruz, 1156 High St., Santa Cruz, CA, USA 95064}
\affil{Kavli Institute for the Physics and Mathematics of the Universe (Kavli IPMU), 5-1-5 Kashiwanoha, Kashiwa, 277-8583, Japan}

\author{Stuart D. Ryder}
\affil{Department of Physics \& Astronomy, Macquarie University, NSW 2109, Australia}
\affil{Macquarie University Research Centre for Astronomy, Astrophysics \& Astrophotonics, Sydney, NSW 2109, Australia}

\author{Marc Rafelski}
\affil{Space Telescope Science Institute, Baltimore, MD 21218, USA}
\affil{Department of Physics \& Astronomy, Johns Hopkins University, Baltimore, MD 21218, USA}

\author{Marcel Neeleman}
\affil{Max-Planck-Institut f\"ur Astronomie, K\"onigstuhl 17, D-69117, Heidelberg, Germany}

\author{Jean-Pierre Macquart}
\affil{International Centre for Radio Astronomy Research, Curtin University, Bentley WA 6102, Australia}

\author{Nicolas Tejos}
\affil{Instituto de F\'isica, Pontificia Universidad Cat\'olica de Valpara\'iso, Casilla 4059, Valpara\'iso, Chile}

\author{Regina A. Jorgenson}
\affil{Maria Mitchell Observatory, 4 Vestal Street, Nantucket, MA 02554, USA}

\author{Cherie K. Day}
\affil{Centre for Astrophysics and Supercomputing, Swinburne University of Technology, Hawthorn, VIC 3122, Australia}
\affil{Australia Telescope National Facility, CSIRO Astronomy and Space Science, PO Box 76, Epping, NSW 1710, Australia}

\author{Lachlan Marnoch}
\affil{Department of Physics \& Astronomy, Macquarie University, NSW 2109, Australia}
\affil{Macquarie University Research Centre for Astronomy, Astrophysics \& Astrophotonics, Sydney, NSW 2109, Australia}
\affil{Australia Telescope National Facility, CSIRO Astronomy and Space Science, PO Box 76, Epping, NSW 1710, Australia}

\author{Shivani Bhandari}
\affil{Australia Telescope National Facility, CSIRO Astronomy and Space Science, PO Box 76, Epping, NSW 1710, Australia}

\author{Adam T. Deller}
\affil{Centre for Astrophysics and Supercomputing, Swinburne University of Technology, Hawthorn, VIC 3122, Australia}

\author{Hao Qiu}
\affil{Sydney Institute for Astronomy, School of Physics, the University of Sydney, NSW 2007, Australia}
\affil{Australia Telescope National Facility, CSIRO Astronomy and Space Science, PO Box 76, Epping, NSW 1710, Australia}

\author{Keith W. Bannister}
\affil{Australia Telescope National Facility, CSIRO Astronomy and Space Science, PO Box 76, Epping, NSW 1710, Australia}

\author{Ryan M. Shannon}
\affil{Centre for Astrophysics and Supercomputing, Swinburne University of Technology, Hawthorn, VIC 3122, Australia}

\author{Kasper E. Heintz}
\affil{Centre for Astrophysics and Cosmology, Science Institute, University of Iceland, Dunhagi 5, 107 Reykjav\'ik, Iceland}

\begin{abstract}
We present a high-resolution analysis of the host galaxy of fast radio burst (FRB)~190608, an SB(r)c galaxy at $z=0.11778$ (hereafter HG~190608), to dissect its local environment and its contributions to the FRB properties. Our Hubble Space Telescope Wide Field Camera 3 ultraviolet and visible light image reveals that the subarcsecond localization of FRB~190608 is coincident with a knot of star-formation ($\Sigma_{\rm SFR} = 1.5 \times 10^{-2}~ \mmsun \, \rm \, yr^{-1} \, kpc^{-2}$) in the northwest spiral arm of HG~190608.  Using H$\beta$ emission present in our Keck Cosmic Web Imager integral field spectrum of the galaxy with a surface brightness of $\mmuhbeta = \sbhbeta$, we infer an extinction-corrected H$\alpha$ surface brightness and compute a dispersion measure (DM) from the interstellar medium of HG~190608 of $\mdmismhost = 94 \pm 38~ \mdmunits$. The galaxy rotates with a circular velocity $v_{\rm circ} = 141 \pm 8~ \mkms$ at an inclination $i_{\mathrm{gas}} = 37 \pm 3^\circ$, giving a dynamical mass $M_{\rm halo}^{\rm dyn} \approx 10^{11.96 \pm 0.08}~\mmsun$. This implies a halo contribution to the DM of $\mdmhalohost = \rm 55\pm25$ \dmunits~subject to assumptions on the density profile and fraction of baryons retained. From the galaxy rotation curve, we infer a bar-induced pattern speed of $\Omega_p=34\pm 6\;\mathrm{km\;s^{-1}\;kpc^{-1}}$ using linear resonance theory. We then calculate the maximum time since star-formation for a progenitor using the furthest distance to the arm's leading edge within the localization, and find $t_{\mathrm{enc}} = 21_{-6}^{+25}$ Myr. Unlike previous high-resolution studies of FRB environments, we find no evidence of disturbed morphology, emission, or kinematics for FRB~190608.

\end{abstract}
\keywords{galaxies: distances and redshifts, spiral arms, star formation, stars: general, radio transient source }

\section{Introduction} \label{sec:intro}

Fast radio bursts (FRBs) are brief ($\delta t \sim $\,ms) pulses of bright ($\gtrsim 1$~Jy\,ms) radio emission detected primarily at meter and decimeter wavelengths \citep{frb_rev1,frb_rev2}. While similar in nature to pulsars and their cousins -- the rotating radio transients (RRATs)—the energetics and stochastic repeating nature of (at least) some FRBs imply a qualitatively distinct physical mechanism \citep{frb_theory}.  Discovered over a decade ago, the frequent localization of these sources is a recent advance enabled by new facilities, operational modes, and extensive follow-up campaigns \citep[e.g.][]{Bannister+19,Ravi+19,Chatterjee+17}. More than seven events with arcsecond or subarcsecond localizations are now associated to their host galaxies, providing first assessments of the population \citep[see][and references therein]{Bhandari+20,Macquart+20,Marcote2020}. 
This first set shows a diversity of galaxy properties, with stellar
masses ranging from $\mmstar \approx 10^{8} - 10^{11} \mmsun$,
specific star formation rates (SFRs) of
$ 10^{-8} \, {\rm yr^{-1}}$
to less than $10^{-11} \, \rm yr^{-1}$, and morphologies
ranging from dwarf to spiral to early-type systems.

As demonstrated by studies of gamma-ray bursts 
\citep[GRBs; e.g.][]{Bloom2002,Prochaska2006}, 
a promising path forward to understanding the origin of 
transient sources is to dissect the 
galaxies that host them, i.e.\ constraining/understanding 
the typical mass, SFR, environment, etc. of galaxies hosting FRBs. While GRBs were early on linked to supernovae (SNe), which pinpointed 
their explosion mechanism, FRBs thus far have no
detected supernova-like optical counterparts \citep{Marnoch+20}. 
Instead, we have to rely on the host and FRB properties for insight
into the progenitor(s).
For GRBs, the former were central to implicating
the collapsar model as the progenitor
of long-duration bursts \citep{Fruchter06}.
This followed from both the association of these GRBs to
star-forming galaxies and that they were typically colocated with the brightest region of UV emission within the galaxy.

Galaxy properties and the local FRB environment may inform 
both progenitor models and other scientific pursuits with FRBs \citep{Eftekhari2017,Tendulkar2017,Prochaska2019,Bhandari+20,Heintz2020}. These include the star-formation rate, the morphology of both the gas and stellar disk, the metallicity, and the location of the event relative to the galaxy
nucleus, its stellar distribution, and its interstellar medium. 
Estimating these properties 
requires deep imaging and multiwavelength
spectra at high spatial resolution 
combined with a high-precision localization. Assessing whether FRBs are produced in specific regions or in distinct small-scale environments will ultimately provide prominent constraints on the progenitor models.

Previous works have examined the environments of two repeating
FRBs with milliarsecond localizations from very long baseline interferometry (VLBI) measurements
\citep[FRBs\,121102 and 1890916.J0158+65;][]{Tendulkar2017,Marcote2020}.  FRB~121102 was identified
with a nebular region in a dwarf, low-metallicity, star-forming galaxy and
is nearly coincident with a persistent radio source.
These associations have inspired and supported FRB models 
related to young, massive stars in metal-poor environments
and to scenarios that invoke active galactic nuclei (AGN) activity \citep{Zhang2019,Katz2017,Vieyro2017}.
In contrast,  FRB 180916.J0158+65 lies in the outer arm
of a more massive spiral galaxy, with an overall low star-formation rate. Repeating FRBs thus appear to not favor any distinct environments (and/or show a broad range of galaxy environments).

In this study, we focus on an apparently nonrepeating event, 
FRB~20190608B, hereafter FRB 190608, discovered and localized with the Australian Square Kilometer Array Pathfinder (ASKAP) \citep{Macquart+20}. 
Its host galaxy, identified as \sdssname\ and hereafter called \hgname, is also a spiral galaxy with high stellar mass 
($\mmstar \approx 10^{10.4} \, \mmsun$)
and a star formation rate of ${\rm SFR} \approx 1.2 \, \mmsun \, \rm yr^{-1}$
\citep{Bhandari+20}.
Full analysis of the FRB baseband data \citep{Day+20},
yields a dispersion
measure (DM) of ${\rm DM_{FRB}} = \dmval~\mdmunits$, a rotation
measure (RM) of ${\rm RM_{FRB}}  = \rmvalue \pm \rmerr \; \rmunits$, and a scattering time of \dt = $\dtval \pm \dterr \;\rm \dtunits$ at 1.28 GHz.
The DM value, when corrected for Galactic contributions, 
exceeds the average cosmic value by nearly a factor of 2
\citep{Macquart+20}, indicating either a large
host contribution and/or an overdensity
of foreground gas \citep{Simha+20}.
Meanwhile, the RM and $\tau$ measurements respectively
indicate a magnetized plasma foreground to FRB~190608 and suggest
propagation through a turbulent medium.
The primary motivation for this study is to examine
the environment of FRB~190608 in the context of 
these propagation effects.

To this end, we have obtained UV imaging of \hgname\ 
with the Hubble Space Telescope (HST) at a spatial
resolution 
$\approx 0.''1$.
These data are complemented by integral field spectroscopic
observations with the Keck Cosmic Web Imager on the
Keck~II telescope.  The resultant data cube maps the
nebular emission lines across the galaxy, albeit at 
seeing-limited ($\approx 0.''9$) resolution.
Together, these data are used to explore and
constrain the physical environment of FRB~190608.

This paper is organized as follows.
We detail the datasets and their reduction 
in Section \ref{sec:data}. 
Section \ref{sec:measure} provides
the primary measurements of the dataset, including  kinematic modeling of the galaxy. 
We then consider how the host galaxy may contribute to the dispersion measure, rotation measure, and scattering observed for the FRB in Section \ref{sec:analysis}. In Section \ref{sec:discussion}, we compute a maximum time for bar-induced star formation for a stellar-type progenitor using linear resonance theory, and put our general findings in the paper in context of the spiral host galaxy of FRB 190916.J0158+65 \citep{Marcote2020}. Finally, we summarize our work in Section \ref{sec:conclusion}. Throughout this work, we have assumed cosmological parameters from the results of \citet{Planck15}.

\begin{figure*}
\centering
\includegraphics[width=0.95\textwidth]{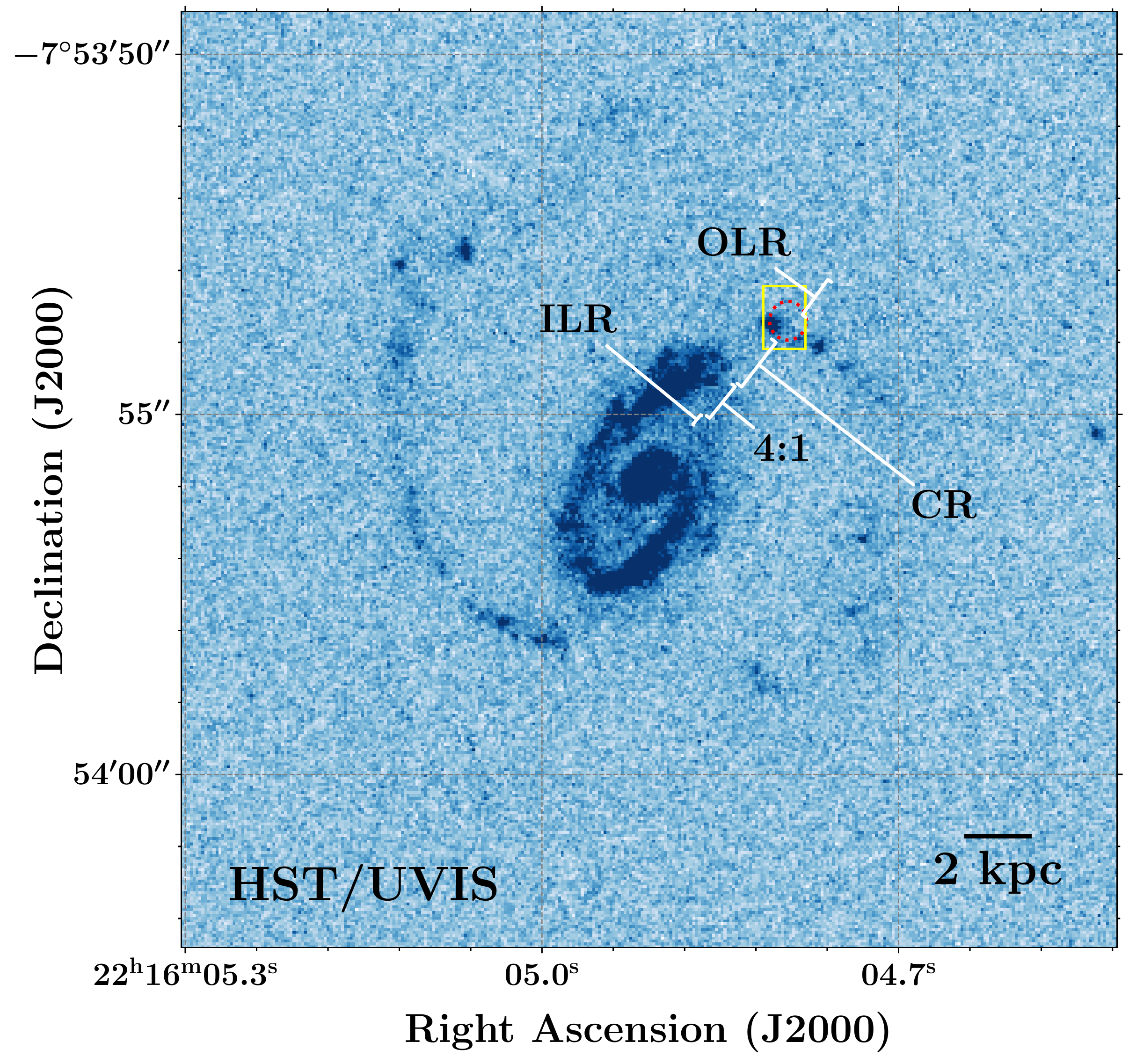}
\caption{The \textit{HST/UVIS} F300X image of \hgname. 
The galaxy shows a bulge, a ring, and prominent spiral characteristics
of an SB(r)c galaxy.
The 1$\sigma$ uncertainty in the FRB position (red oval)
is coincident with a star-formation region in the northwest spiral arm. We estimate a surface brightness of
$\mu_{\rm UV} \approx \mathrm{(4.65\pm0.18)\times10^{-3}\;mJy\;arcsec^{-2}}$ within the red ellipse.
The yellow box represents the region evaluated in the Keck Cosmic Web Imager (KCWI) analysis. The positions along the major kinematic axis for the inferred Lindblad resonances (see Section \ref{sec:pattern}) are denoted by the white lines.}
\label{fig:hst}
\end{figure*}

\section{Data} \label{sec:data}

\subsection{ASKAP}

At Coordinated Universal Time (UT) 22:48:12 on 2019 June 08, FRB~190608 was detected by the Commensal Real-time ASKAP Fast Transients (CRAFT) survey on the Australian Square Kilometer Array Pathfinder (ASKAP) and was subsequently localized to ${\rm \rafrb}, \, {\rm \decfrb}$  (right ascension, declination, J2000).

The high-time resolution analysis in \citet{Day+20} yielded a more precise FRB localization than reported in \citet{Macquart+20}. The statistical uncertainty in the position is described by an ellipse with $\sigma_{\rm RA,Stat} = \stataerr''$ for one axis and $\sigma_{\rm Dec,Stat} = \statberr''$ for the other. In addition, the registration of the ASKAP image in the International Celestial Reference Frame is subject to an uncertainty determined by the number and brightness of background radio sources present in the image. This systematic uncertainty is estimated to be an ellipse with $\sigma_{\rm RA,Sys} = \sysaerr''$ and $\sigma_{\rm Dec,Sys} = \sysberr''$. The errors along respective axes are added in quadrature for a final FRB position uncertainty (68\%\ c.l.) with a semi-major axis (RA) of $\sigma_{\rm RA} \approx 0.26''$ and a semi-minor axis (Dec) of $\sigma_{\rm DEC} \approx 0.25''$.

The dispersion measure of FRB~190608 is measured
at $\mdmfrb = \dmval \, \mdmunits$ \citep{Day+20},
well exceeding the Galactic interstellar medium (ISM) estimate along its
sightline \citep[$\mdmmwism \approx 33\ \mdmunits$;][]{Cordes}.

Analysis of the baseband
data reveals a large rotation measure \citep{Day+20}: 
$\rm RM = \rmvalue \pm \rmerr \, \rmunits$.  
Furthermore, the observed pulse is broad ($\approx \dtval$\,ms at 1.28~GHz)
and shows a roughly $\nu^{-4}$ dependence.
For the following, we assume a scatter broadening
of $\approx 2$\,ms \citep{Day+20}.
These measurements respectively
indicate propagation through magnetized and turbulent plasma.


\begin{figure*}
\centering
\includegraphics[width=0.95\textwidth]{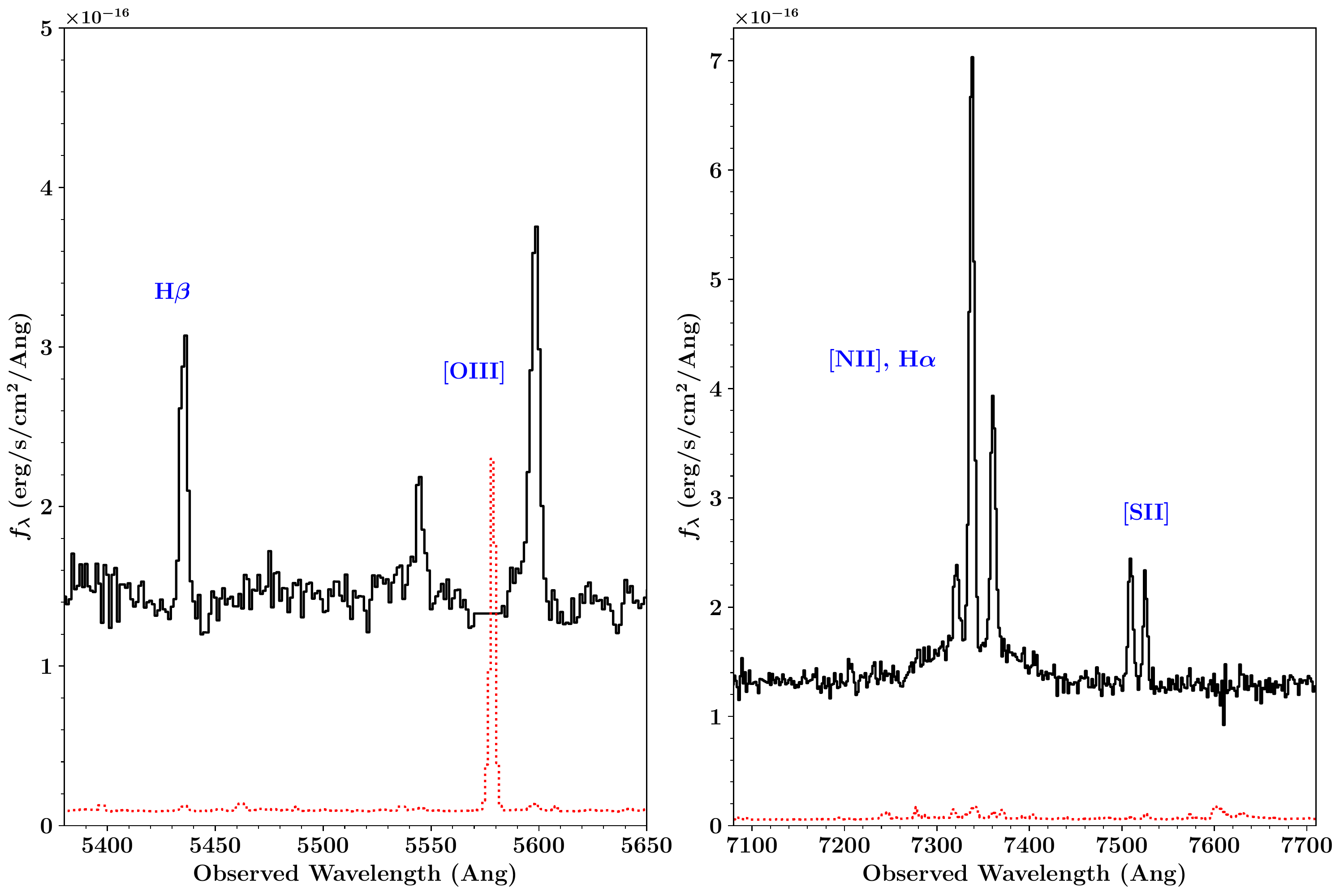}
\caption{Portions of the $3''$ diameter aperture SDSS spectrum of \hgname\ centered
on standard nebular emission lines. The red dotted line is the error ($\sigma$) of the spectrum. Note the broad
H$\alpha$ emission underneath the narrow H$\alpha$ and [N\,{\sc ii}]
emission are hallmarks of a Type~1 AGN.}

\label{fig:sdss_spectrum}
\end{figure*}

\subsection{HST Ultraviolet and Visible Light Observations and Reduction}

On UT 2019 October 11, we obtained near-ultraviolet imaging with the Wide Field Camera 3 (WFC3) on board the HST. We used the wide F300X filter, covering rest frame $\sim2400-3200$ \AA. While this filter has a red tail out to $\sim4000$ \AA, it has a high UV throughput of $\sim15$\% with a wide passband. The single orbit was divided into 4$\times$600 second exposures. To minimize the effects from charge transfer degradation and maximize UV sensitivity, 
the target was placed near the readout on chip 2 on amplifier C, and the exposures include 9e$^-$ postflash per exposure to reach a 12e$^-$ per pixel background.  The images are dithered with a box dither pattern 5 times larger than the standard WFC3 ultraviolet and visible light (UVIS) box dither pattern to minimize residual background patterns. 

The four images are calibrated with custom processing similar to that described in \cite{Rafelski2018}, which will be described in more detail in \citet{Prichard+21}. In short, we use a new correction for the charge transfer efficiency, use concurrent dark exposures for superdark creation,  equalize the number of hot pixels detected as a function of the distance to the readout, and normalize the amplifiers to each other. The near-UV images are combined using \verb|AstroDrizzle| \citep{Avila2015} at their native 40 mas plate scale with a pixel fraction of 0.8, include sky subtraction, and are oriented North up and East left. The images are aligned to GAIA DR2 \citep{Gaia2018} using \verb|TweakReg| \citep{Avila2015} and have an astrometric accuracy of $\approx$0.02$''$.
Figure~\ref{fig:hst} presents the combined image and the FRB localization.

\subsection{SDSS}

The localization of FRB~190608 associates it 
to a galaxy cataloged by the Sloan Digital Sky Survey (SDSS), \sdssname.
The SDSS spectrum recorded from the inner $3''$-diameter 
of the galaxy yields a redshift $z=\zhg$. 
Figure~\ref{fig:sdss_spectrum} shows a portion of the SDSS 
spectrum focusing on a series of standard nebular lines.
In addition to the narrow nebular emission characteristic
of star-forming regions, the data also show broad
H$\alpha$ emission indicative of a Type~1 AGN 
\citep{Stern2012a}.
\cite{Bhandari+20} further analyzed this spectrum and
the SDSS photometry to estimate
a stellar mass of $\mmstar \approx 10^{10.4}\,\mmsun$,
SFR~$\approx 1.2\,\mmsun \, \rm yr^{-1}$,
and metallicity of $Z = 0.009$.

\subsection{Keck/KCWI Observations and Reduction}

On UT 2019 September 30 and October 01, we obtained a combined set of 3$\times$900 s exposures of \hgname\ with the KCWI on the Keck~II telescope. The data were obtained with the integral-field unit (IFU) in the ``Medium" slicer position with the ``BM" grating, resulting in a field of view (FOV)
of $16.5'' \times 20.4''$ and a spectral resolution of R=5000 (FWHM). 
Both observing nights were clear with seeing of FWHM$\sim0.9''$.

The data were processed with the standard KCWI Data Reduction Pipeline\footnote{\url{https://github.com/Keck-DataReductionPipelines/KcwiDRP}} \citep{Morrissey2018}. Flat field and arc calibrations were made using data from the September 30th run. Sky subtraction sampling was limited to slices beyond the host galaxy to avoid subtracting any signal in the spectra. The pipeline also corrected for differential atmospheric refraction across the FOV. Finally, we flux calibrate the spectra with standard star observations of BD+25 4655 and G191B2B with the same configuration as for the host galaxy observations.

The three exposures were aligned and combined
to increase the signal-to-noise ratio using the \verb|CWITools|\footnote{CWITools: \url{https://github.com/dbosul/CWITools}} package \citep{OSullivan2020,CWITools}, and were rebinned so both spatial axes are in the same scale. Each spaxel covers a projected size of $0.29''\times0.29''$. We then converted the wavelengths from air to vacuum and applied a geomotion correction based on the sightline and time of observation. Given the absence of any other bright source in the KCWI FoV, we have forced the astrometric solution of \hgname\ from SDSS to the centroid of the KCWI data identified with intensity contours and estimated an uncertainty of the order of half a spaxel ($\sim 0.15''$).


\begin{figure}
\centering
\includegraphics[width=0.49\textwidth]{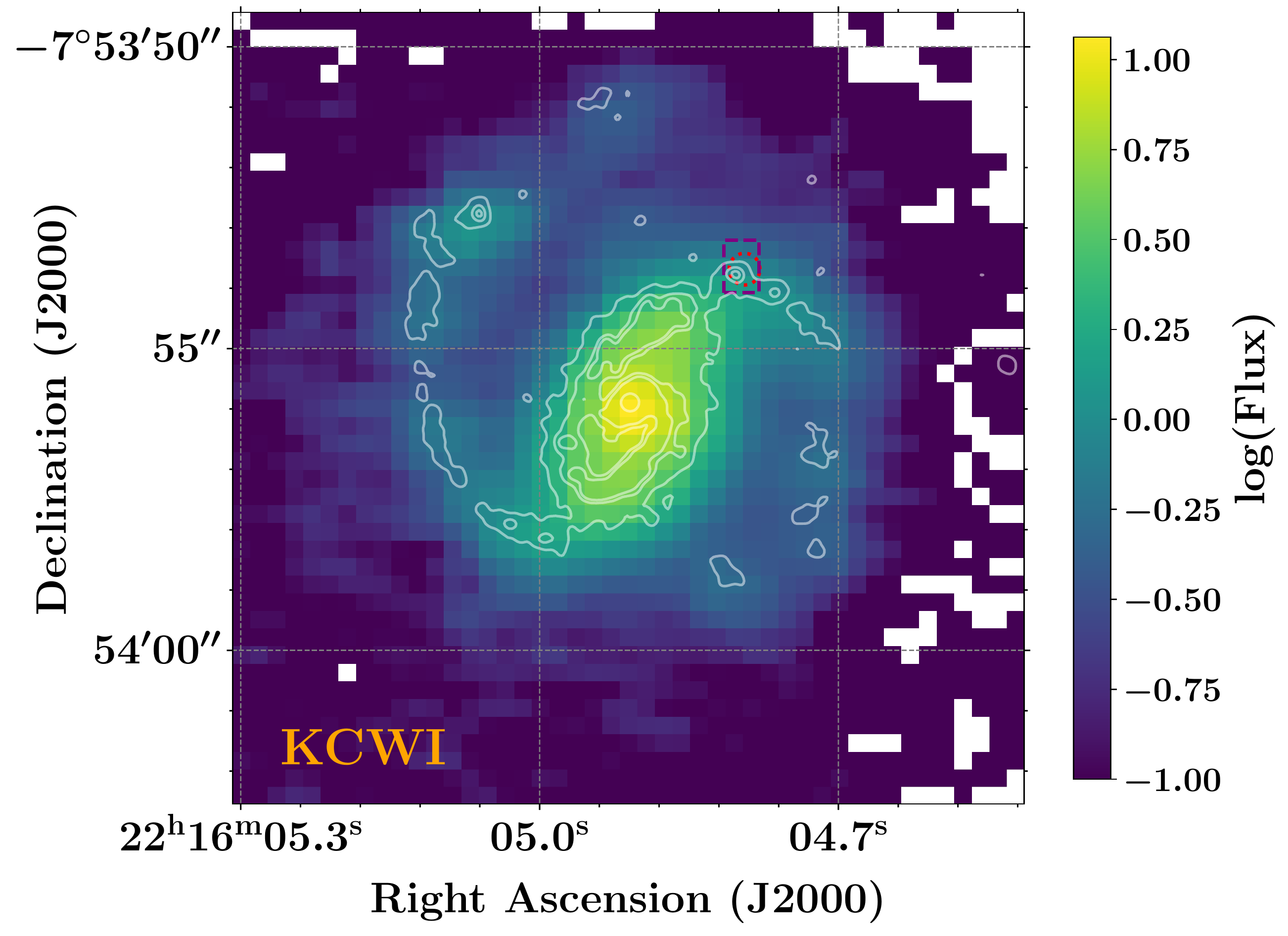}
\caption{A psuedo narrowband H$\beta$ image of \hgname\ created by summing the KCWI spaxels between $\lambda_{\rm obs} \approx 5425.4-5445.4 $\AA. Linear contours from the Gaussian-smoothed \hst\ data are overlaid in white. The 1-$\sigma$ uncertainty in the FRB position is overlaid in red, and the 
6~spaxel box used for flux measurements is 
shown by the dashed purple rectangle.}
\label{fig:Hbeta}
\end{figure}

\section{Measurements} \label{sec:measure}

\subsection{HST/UVIS Imaging}\label{sec:hstflux}
In Figure~\ref{fig:hst}, we present the processed \hst\textit{/UVIS} image with the FRB localization indicated by the red circle. 
Its centroid lies just off of bright
UV emission associated with the northwest spiral arm, which is 
encompassed by the 68\%\ position uncertainty. 
We classify the host as a grand design SB(r)c galaxy based on the bright ring of star formation connecting the relatively small bulge to the two spiral arms. The ring encircles the extent of the bar, and due to the galaxy's inclination to our line of sight coupled with the major axis alignment of the bar, it then appears to be elliptical even though if viewed face-on it would appear circular (unlike the bar itself).

The UV emission traces massive star formation, which highlights the leading edge of the bar and/or inner ring. Bars in general are indistinct from the bulge and inner disk in their stellar kinematics, though the gas kinematics may display evidence of bar-driven gas inflow that feeds star formation/AGN (not seen at this resolution). The bar itself is more evident in the \textit{g}-band X-shooter image of the host in Figure 1 of \citet{Bhandari+20}. We attribute the tightly-wound inner structure to star formation near the second harmonic (4:1) Lindblad resonance with the bar \citep{Lindblad1959,EE1989,EE1990,Ryder1996,EE1996}. Star-forming rings are known to be clumpy \citep[e.g., ``pearls on a string", as studied by][]{Boker2008}, and tend to fade azimuthally due to aging as HII regions drift downstream from where the arms channel gas down on to the ring \citep{Ryder2001}. These two effects may explain the clumpiness and asymmetry observed in Figure \ref{fig:hst}.

We estimate the galaxy center from the flux-weighted centroid
of the UV flux within the bulge and find a position of
22h16m04.90s, $-$07d53m55.91s (R.A, decl)
with a statistical uncertainty of $0.08''$. Two methods of determining the centroid (\verb|photutils.centroids.centroid_com| and \verb|photutils.centroids.centroid_1dg|) agree to within 1 pixel ($0.04''$), providing our estimate of the
systematic uncertainty.
This yields a projected offset of $2.93'' \pm 0.28''$
for FRB~190608 from the center of the galaxy with the uncertainty dominated by the
systematic error of the FRB localization.
At $z=\zhg$, this corresponds to a projected physical offset of 
$R_{\perp} = 6.44 \pm 0.62$ kpc, consistent with 
the estimate of \cite{Bhandari+20}.

Within the FRB positional uncertainty (see red oval in Figure \ref{fig:hst}), we measure an integrated UV electron flux (number of electrons accumulated in the detector) of $\rm{1.457 \pm 0.050\;e^-\;s^{-1}}$. The uncertainty is determined by taking the inverse square root of the inverse sensitivity map, a data product from the HST/UVIS pipeline.
Adopting the AB magnitude zero point 
($ZP_{\rm F300X}=25.069$) for the F300X filter 
\citep{wfc3_handbook}, 
we determine a UV flux of $f_{\nu,UV} = \mathrm{(3.09 \pm 0.12)\times 10^{-4}\;mJy}$ ($m_{\rm AB} = 25.2$) and subsequently a UV surface brightness of $\mu_{\rm UV} = \mathrm{(4.52\pm0.18)\times10^{-3}\;mJy\;arcsec^{-2}}$.


\begin{figure*}
\centering
\includegraphics[width=0.95\textwidth]{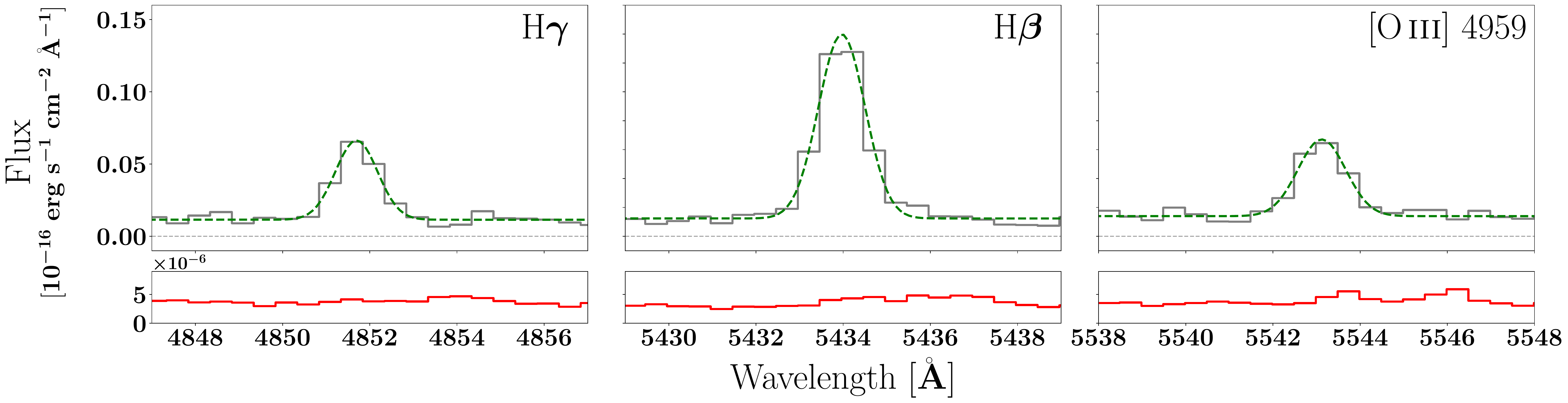}
\caption{Summed spectra of the 2$\times$3 spaxel box at the FRB position for each identified emission line. The gray solid lines are the spectra, the green dashed lines are the fitted Gaussians to each line, the dotted gray lines show the zero-level flux, and the solid red lines are the variance.}
\label{fig:spectra}
\end{figure*}


\begin{figure}
\centering
\includegraphics[width=0.49\textwidth]{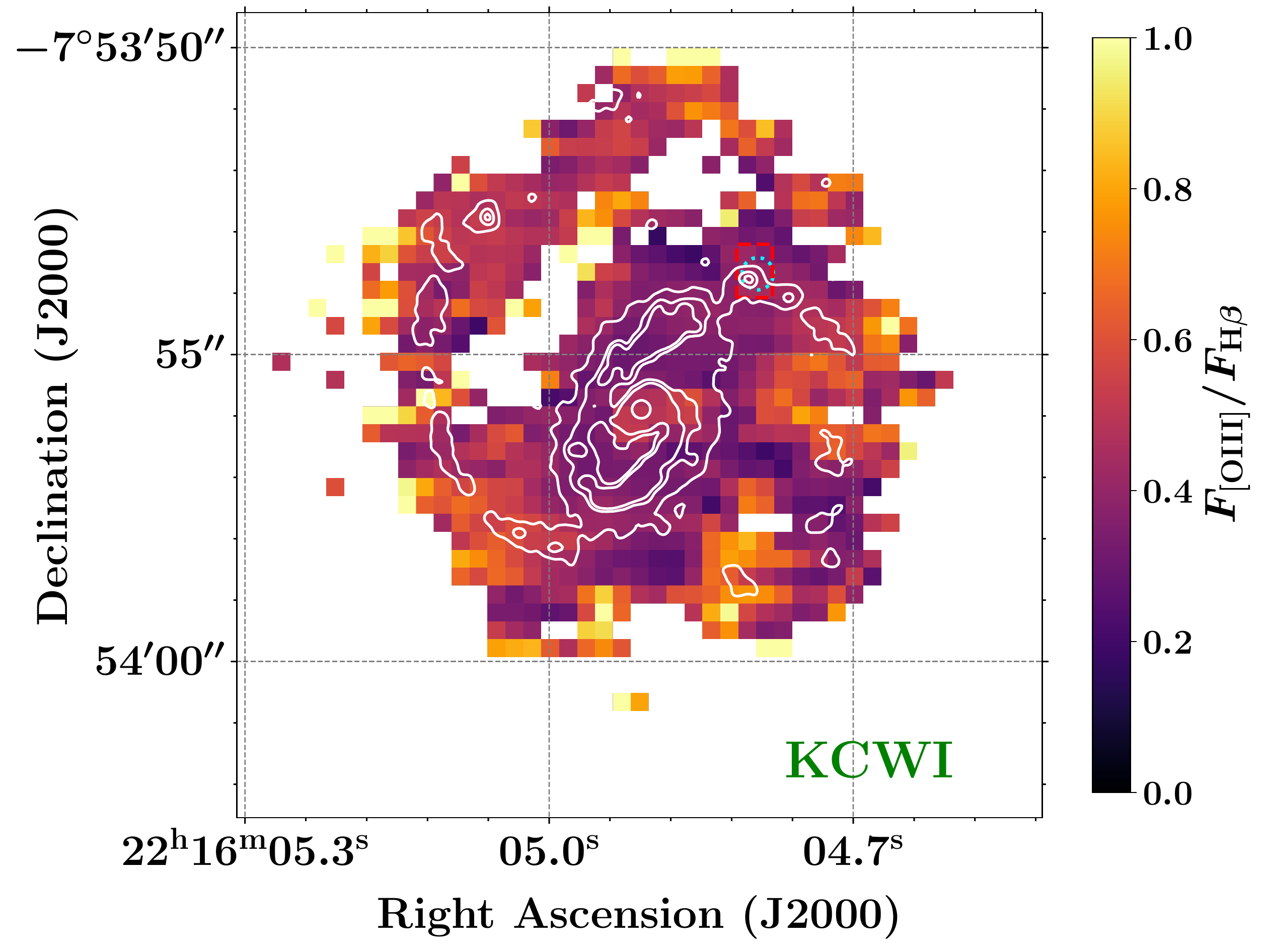}
\caption{Map of the ratio of [O\,{\sc iii}] and H$\beta$ flux. Each emission line above a 3-$\sigma$ detection was fitted with a Gaussian to determine the flux. 
Like Figure~\ref{fig:Hbeta}, contours from the \hst\ data are overlaid in white, the dashed red rectangle is a box used for measurements regarding the FRB position, and the cyan dashed circle is the 1$\sigma$ uncertainty in the FRB position.}
\label{fig:ratio}
\end{figure}


\subsection{Keck/KCWI Line Fluxes} \label{sec:kcwiflux}

\begin{table*}
\centering
\caption{HG 190608 Emission Line Measurements} \label{tab:linefits}
\begin{tabular}{cccc}
\tablewidth{0pt}
\hline
\hline
&&KCWI FRB Position\tablenote{Measured from the 6 spaxels encompassing the FRB position (See Figure \ref{fig:Hbeta})}& SDSS\tablenote{Measurements taken from \cite{Bhandari+20}} \\
\hline
Line & $\lambda_{\rm rest}$\tablenote{Vacuum wavelengths adopted from \url{http://classic.sdss.org/dr6/algorithms/linestable.html}} & Line Flux & Line Flux \\
&$(\mathrm{\AA})$&($\mathrm{10^{-17}\;erg\;s^{-1}\;cm^{-2}}$)&($\mathrm{10^{-17}\;erg\;s^{-1}\;cm^{-2}}$)\\
\hline
H$\gamma$ & 4341.68 & 0.67 $\pm$ 0.06 & 39.2 $\pm$ 3.4 \\
H$\beta$ & 4862.68 & 1.71 $\pm$ 0.11 & 83.7 $\pm$ 3.3\\
$\lbrack$O\,{\sc iii}] 4959 & 4960.295 & 0.73 $\pm$ 0.05 & $50 \pm 2$\\
\ha & 6564.61 & -- & 277 $\pm$ 4.15\\
\hline
\end{tabular}
\end{table*}


Complementing the \hst\ observations, the KCWI data
cube yields measurements on the nebular emission lines
within the rest-frame interval of $\lambda_{\rm rest} \approx 4723-5577$\AA.
Figure~\ref{fig:Hbeta} shows a pseudo narrowband image of HG~190608 centered on H$\beta$ emission ($\lambda_{\rm obs} \approx 5425.4-5445.4$\AA). 
Overlaid on the H$\beta$ psuedo image are the contours of
UV emission from the \hst\ image.
As expected, there is a close correspondence between the two;
the KCWI data are effectively a seeing-smoothed description
of the star-forming regions.

We define a 6~spaxel region encompassing the $1\sigma$ 
localization circle of FRB~190608 
(see the purple box in Figure~\ref{fig:Hbeta})
from which we extract a 1D spectrum to
analyze the H$\gamma$, H$\beta$, and [O\,{\sc iii}]~4959
emission lines. The [O\,{\sc iii}]~5007 line was beyond the acceptable wavelength range due to our instrument configuration.
We measured line fluxes
by fitting a Gaussian to each line and
estimate uncertainties by subtracting the Gaussian fits from the data and then calculating the standard deviation of the residual spectrum. This method yielded uncertainties about an order of magnitude higher than the formal errors from the fit.
Figure~\ref{fig:spectra} shows the emission lines and fits,
and Table~\ref{tab:linefits} reports the measurements.

From the ratio of H$\gamma$ to H$\beta$ flux at the FRB position, 
we can estimate the internal reddening from dust in the galaxy.
Comparing the observed ratio 0.391 to the
theoretical value of 0.466 from \cite{Osterbrock2006} 
and adopting the extinction curve from \cite{Cardelli1989},
we estimate $A_V = 1.28$\,mag.
This is comparable to the extinction estimate of 
\cite{Bhandari+20} from their analysis of the SDSS spectrum from the inner region.

From the H$\beta$ line flux and the area of the
integration region, 
we determine an average H$\beta$ surface brightness 
of $\mmuhbeta = \sbhbeta$, uncorrected for dust. 
We may also estimate the dust-corrected
H$\alpha$ surface brightness \muchalpha, as follows.
We adopt the observed H$\beta$ surface brightness 
and the estimated internal extinction $A_V$.
We assume the intrinsic $f_{\rm H{\alpha}}/f_{\rm H{\beta}}$ follows
the putative ratio of 2.87 for H\,{\sc ii} regions \citep{Osterbrock2006} 
and find that $f_{\rm H{\alpha}}/f_{\rm H{\beta}}$ = 4.63.
This yields $\mmuchalpha = \sbchalpha$.

The measured line fluxes at the FRB location 
show a lower [O\,{\sc iii}]/H$\beta$ ratio than through the
SDSS fiber on the central regions, which may be explained by the presence of AGN emission in the nucleus (as also indicated by the broad \ha\ emission component).
To explore this further, we generated a flux-ratio image by fitting each [O\,{\sc iii}] and H$\beta$ emission line with a Gaussian and integrating to determine the flux (rather than direct integration, which produced a noisier result). Only spaxels that had a signal-to-noise ratio $>$ 3 for both H$\beta$ and [O\,{\sc iii}] are plotted in Figure~\ref{fig:ratio}. We find that the central region of the galaxy has an [O\,{\sc iii}]/H$\beta$ ratio of $\approx 0.5$ while that of the 
FRB region is $\approx  0.4$.

\begin{figure*}
\centering
   \includegraphics[width=0.49\textwidth]{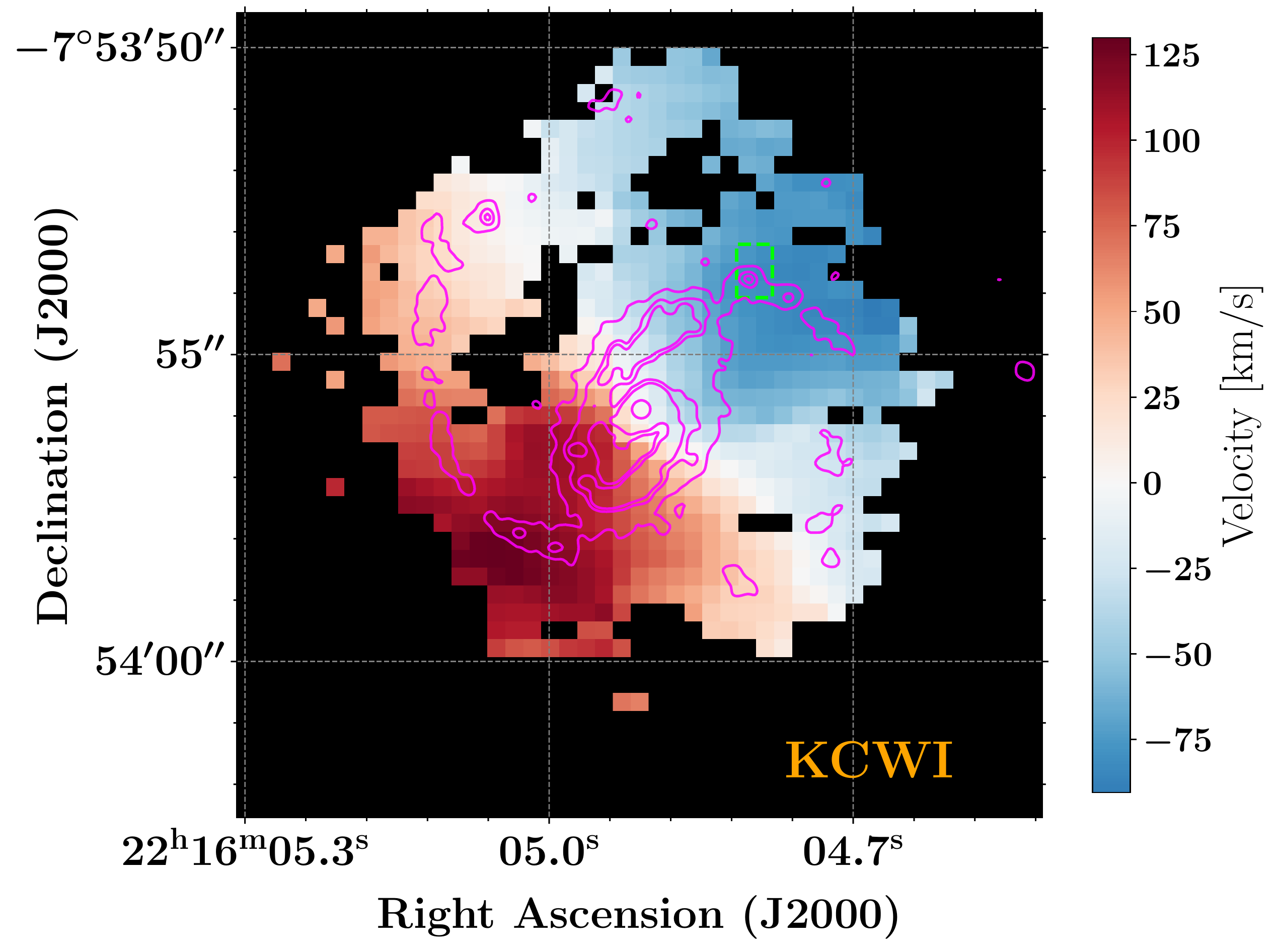} \label{fig:vel}
   \includegraphics[width=0.49\textwidth]{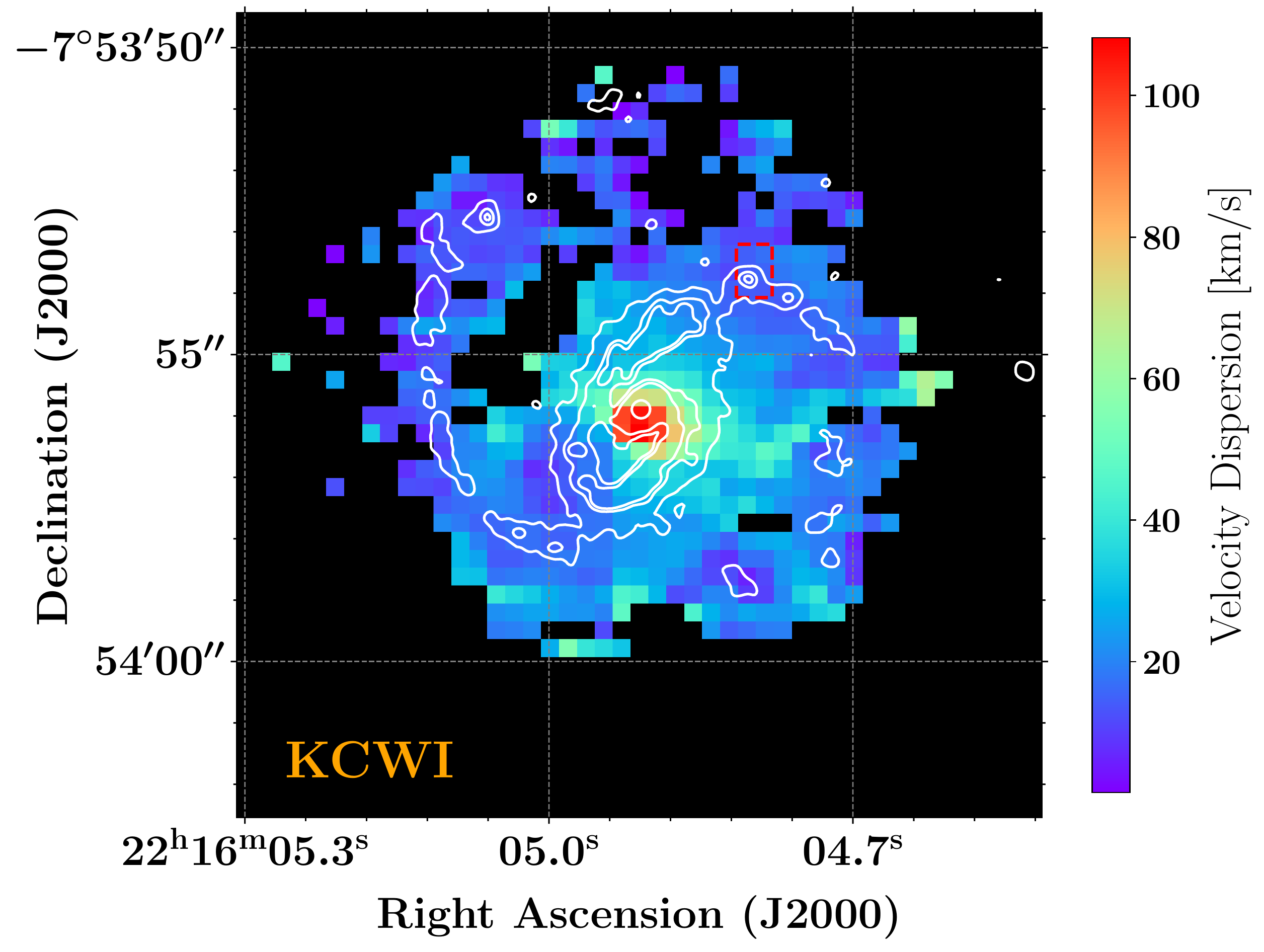}\label{fig:disp}
\caption{Left: A Doppler velocity map of the KCWI data with the zero-point set by the SDSS redshift. Spaxels for which we have at least a 3-$\sigma$ detection of the  H$\beta$ line are fit with a Gaussian. We adopt the central wavelength of the fit and compute the velocity for each spaxel. We overlaid the contours from the \hst\ data in magenta, and the dashed green rectangle is the 2$\times$3 spaxel box, which we averaged to compute the velocity at the FRB position.
Right: A velocity dispersion map of the KCWI data showing only spaxels with a 3$\sigma$ detection of H$\beta$. We subtracted the dispersion of $25.5\  \mkms$ resulting from the resolution of the instrument configuration \citep{Morrissey2018}. The \hst\ contours are in white, and the dashedred rectangle is the region over which we computed the average dispersion for the FRB position.}
\label{fig:kin}
\end{figure*}


\subsection{Keck/KCWI Kinematics} \label{sec:kin}

In Figure~\ref{fig:kin}, we present a map of the ionized gas 
kinematics of \hgname\ 
-- the velocity $\delta v$ relative to 
$z=\zhg$ and the RMS velocity dispersion $\sigma$ --
as measured from the KCWI data. 
To determine the gas velocities, we used a Gaussian model to fit the H$\beta$ lines detected at each spaxel, with the rest wavelength determined by the redshift. 
The data reveal rotation characteristics of a disk galaxy,
with the region presumed to host the FRB - in the northwest spiral arm - approaching us while the southeast spiral arm is receding. 
The FRB appears to lie on the kinematic major axis of the galaxy, meaning that it lies between the ``near" and ``far" sides of the galactic disk. However, the prominent inner region revealed in the HST/UVIS data in Figure \ref{fig:hst} consists of a tight spiral or ``ring'' that may be associated with gas compression and star formation driven by the second harmonic (4:1) Lindblad resonance. The inclination of the bulge, the spiral arms, and the outer stellar disk may all be different, suggesting a warped disk. We explore this further in Section \ref{sec:model}.

To calculate the velocity dispersion, we subtract off the instrumental velocity dispersion of $25.5\ \mkms$ in quadrature. We observe a sharp peak in dispersion at the center of the galaxy of $\approx$ 108 km $\rm s^{-1}$, consistent with the dispersion measured in the SDSS spectrum (110.85 $\pm$ 11.492 km $\rm s^{-1}$), that falls more quickly along the major axis than the minor axis.

Averaging the velocity and velocity dispersion images over the
6~spaxel box covering the  FRB uncertainty region, we estimate
$\delta v = \dv$ (relative to $z=\zhg$)
and $\sigma = \vsigma$. Reported uncertainties were determined by the Gaussian fit parameters for the H$\beta$ emission line. At the FRB position, we observe no peculiar velocity behavior when compared to a similar position along the southeast spiral arm.

\vskip 0.1in

\subsubsection{Spiral Galaxy Model} \label{sec:model}
To estimate the kinematical parameters for HG~190608, such as its rotational velocity and intrinsic velocity dispersion, we model the H$\beta$ emission line data using the Python-based code \texttt{qubefit}\footnote{\url{https://github.com/mneeleman/qubefit}} \citep{Neeleman2019}. This code fits the continuum-subtracted H$\beta$ data cube to a model data cube generated from a user-defined model, which is convolved with the point spread function and line spread function of the instrument. This 3D approach minimizes biases in the kinematical properties of the galaxy caused by the finite resolution of the instrument and observations \citep[e.g.][]{DiTeodoro2015}. 
The best-fit parameters and associated uncertainties are determined by the code through a Markov Chain Monte Carlo approach, whereby the parameter space is sampled using an affine-invariant sampler, \texttt{emcee} \citep{Foreman-Mackey2013}. Flat priors are assumed for all parameters.

As HG~190608 shows two distinct spiral arms, we took the thin and thick disk galaxy models described in \citet{Neeleman2019} and added in a spiral density wave described by a Gaussian profile in the azimuthal direction whose central position, $\psi_{\rm c}$, varies with radius, $r$, according to: $\psi_{\rm c}(r) = \psi_{\rm c, 0} + k*r$, where $k$ describes the tightness of the spiral structure and $\psi_{\rm c, 0}$ the position of the spiral structure at the center.
Finally, the H$\beta$ intensity of the spiral structure was taken to obey a simplified step function with a constant value for the intensity, $I_{\rm s}$, below the cutoff radius of the spiral, $r_{\rm s}$. This spiral model adds four additional parameters to  the thin and thick disk models.

We ran fits for both the thin and thick disk models using 300 walkers and 1000 runs each. The results of the fitting are given in Table \ref{tab:modres}. The best-fit models yield an inclination angle of $i \approx \iangle$, where we have adopted the average result of the two models and an uncertainty that is inclusive of both results. We also find a circular rotation speed corrected for the inclination of $v_{\rm circ} = \vcirc$ between the two models. 

\begin{figure}
\centering
   \includegraphics[width=0.47\textwidth]{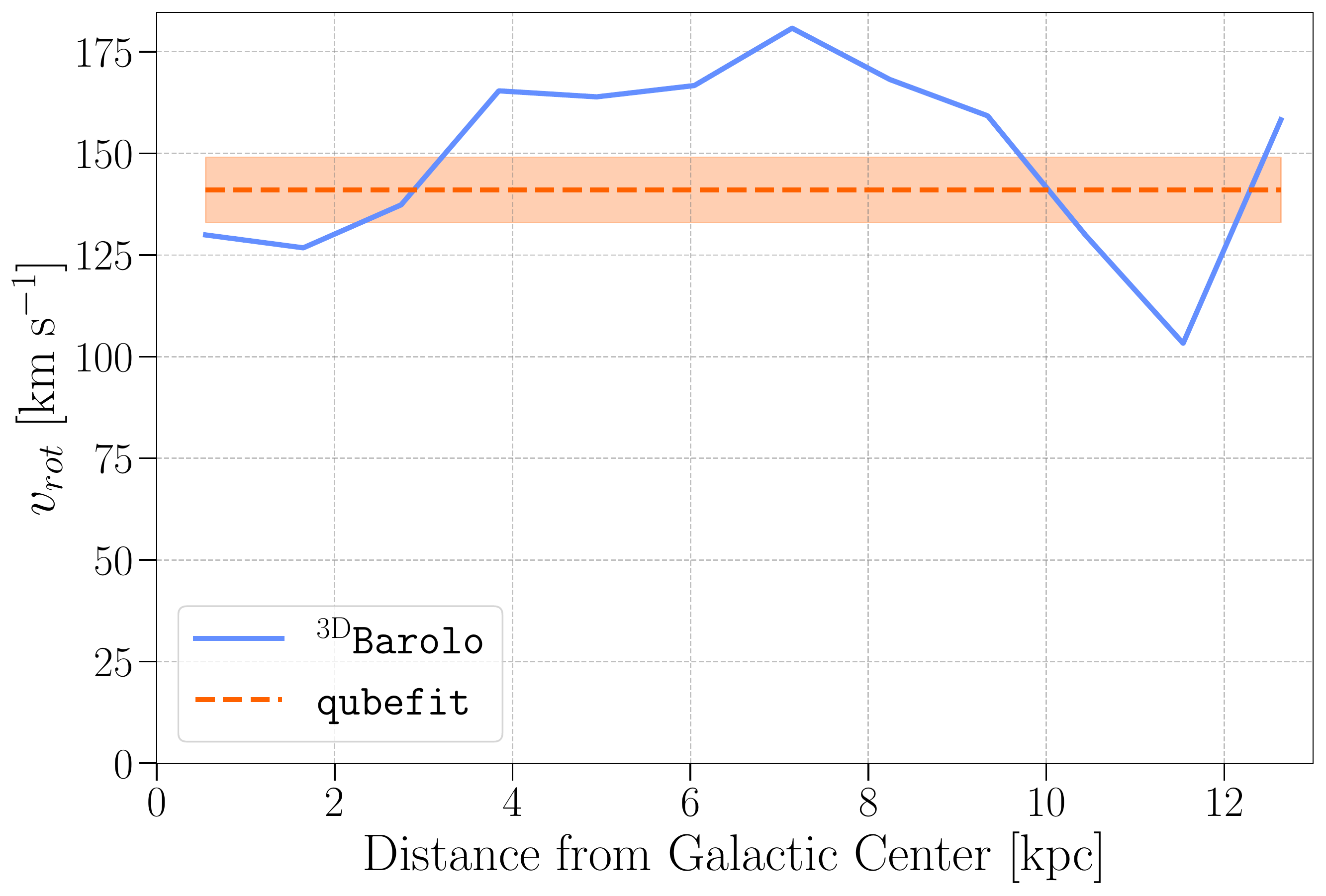}
\caption{Inclination-corrected velocity rotation curves produced with \texttt{$^{3{\rm D}}$Barolo} in blue and \texttt{qubefit} in orange. The shaded orange region represents the 1$\sigma$ uncertainty from the constant velocity fit from \texttt{qubefit}.}
\label{fig:vrot}
\end{figure}

These values are in good agreement with the results obtained using the 3D titled-ring fitting routine, \texttt{$^{3{\rm D}}$Barolo} \citep{DiTeodoro2015}. This code uses a series of concentric rings to estimate the kinematics of each ring and is therefore sensitive to potential warps of the gas disk. However, no such warps were observed in the ionized gas of HG~190608 in a model fit with 12 rings. Using the pixel-by-pixel normalization in \texttt{$^{3{\rm D}}$Barolo}, we determine an inclination and circular rotation speed of 37$\degr$ and 165~$\mkms$, respectively. The rotation curves produced by both fitting routines are presented in Figure \ref{fig:vrot}.

\begin{figure*}
\centering
   \includegraphics[width=0.95\textwidth]{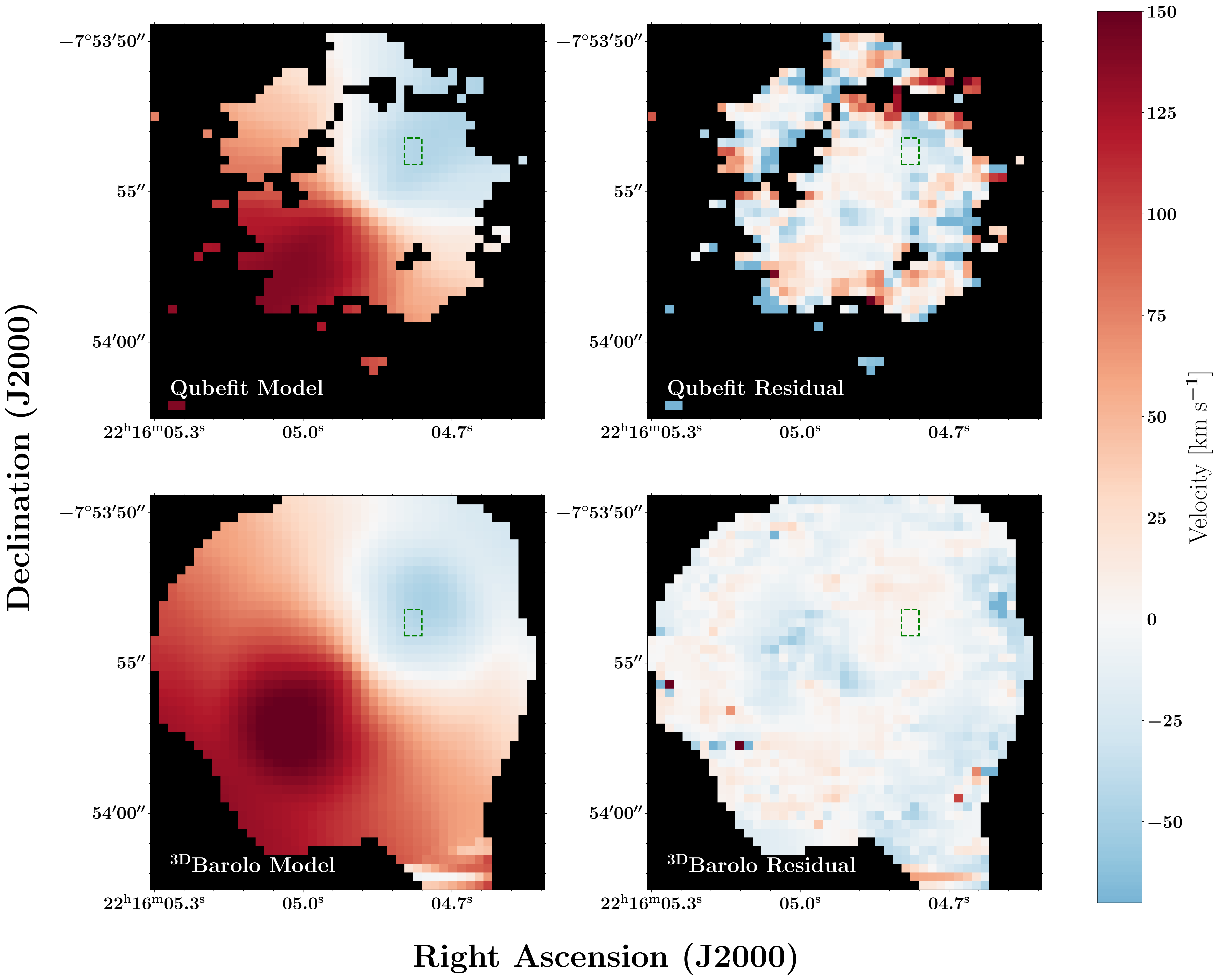}
\caption{Moment-1 (velocity) images of the models (left) and residuals (right) derived from \texttt{qubefit} and \texttt{$^{3{\rm D}}$Barolo}. Note that the two programs use different masks (the black pixels) over the spaxels before fitting. The dashed green box represents the FRB position as in previous figures.}
\label{fig:residuals}
\end{figure*}

In Figure \ref{fig:residuals} we plot the model Moment-1 (velocity) images and residuals produced with \texttt{qubefit} and \texttt{$^{3{\rm D}}$Barolo}. The overall character of the models are in good agreement, with the more complex spiral model from \texttt{qubefit} producing more of a mismatch near the outer parts of the masked galaxy. Such residuals can be used to identify gas inflow and outflow toward the center of the galaxy as well as streaming motions along the spiral arms, the bar, and dust lanes. However, we do not identify any consistent residuals between the models that would be indicative of such effects, possibly due to a combination of the spatial resolution and the more face-on inclination.


The inferred $v_{\rm circ} = \vcirc$ allows us to estimate a dynamical mass for the dark matter halo based on the simple model of \citet{Mo1998}, as $M_{halo}^{\rm dyn} \approx 0.1 G^{-1}H^{-1}(z)v^3_{\rm circ}$, where $G$ is the gravitational constant, and $H(z)$ is the Hubble parameter at redshift $z$. We obtain $M_{halo}^{\rm dyn} \approx 10^{11.96 \pm 0.08}\mmsun$, which is consistent with the inferred halo mass from abundance matching (see Section \ref{sec:dm}).


\begin{table*}
\centering
\caption{Kinematic Modeling Results} \label{tab:modres}
\begin{tabular}{cccc}
\tablewidth{0pt}
\hline
\hline
Parameter & Units & Thin Disk Model & Thick Disk Model \\ 
\hline
\textit{$i_{\mathrm{gas}}$}\tablenote{Inclination.} & ($^{\circ}$) & 39.19$^{+0.62}_{-0.49}$ &35.70$^{+0.17}_{-0.17}$ \\
$\alpha$\tablenote{Position angle.} & ($^{\circ}$) & 133.05$^{+0.95}_{-0.33}$ & 135.55$^{+0.15}_{-1.77}$\\
$I_0$\tablenote{Central specific flux per PSF area.} &    ($\mathrm{10^{-16}\;erg\;s^{-1}\;cm^{-2}\;\AA^{-1}\;arcsec^{-2}}$) &159.91$^{+4.70}_{-11.46}$ &143.37$^{+86.49}_{-36.38}$\\
$R_d$\tablenote{Exponential scale length.} & (kpc) &1.98$^{+0.25}_{-0.14}$ &2.13$^{+0.10}_{-0.13}$\\
$Z_f$\tablenote{Thickness of disk; relevant only for the Thick Disk Model.} & (kpc) & - & 0.07$^{+0.02}_{-0.04}$\\
$v_{\mathrm{circ}}$\tablenote{Circular velocity.} & ($\mathrm{km\;s^{-1}}$) &146.14$^{+2.24}_{-4.96}$ & 140.39$^{+6.11}_{-7.02}$\\
$\sigma_v$\tablenote{Velocity dispersion.} & ($\mathrm{km\;s^{-1}}$) & 34.62 $^{+1.09}_{-1.20}$ &38.16$^{+1.61}_{-2.29}$\\
$\phi_{c,0}$\tablenote{Central position of spiral structure.} & ($^{\circ}$) & 110.38 $^{+1.24}_{-1.37}$ & 103.28$^{+5.45}_{-1.28}$\\
$k$\tablenote{Tightness of spiral structure.} & - & -0.20$^{+0.00}_{-0.00}$ & -0.20$^{+0.00}_{-0.00}$\\
$D_\phi$\tablenote{Thickness of spiral arms.} & ($^{\circ}$) & 1.81$^{+1.13}_{-1.16}$ & 3.08$^{+1.56}_{-0.38}$\\
$I_s$\tablenote{Fraction of central intensity for spiral structure.} & - & 1.35$^{+1.72}_{-0.67}$ & 0.74$^{+0.65}_{-0.25}$\\ $r_s$\tablenote{Cutoff radius of spiral 
arms.} & (kpc)& 12.00$^{+0.73}_{-0.47}$& 11.63$^{+0.31}_{-0.13}$\\
\hline
\end{tabular}
\end{table*}


\subsection{Optical Surface Photometric Analysis} \label{sec:ospa}

Figure~\ref{fig:kin} highlights how dependent the kinematical analysis presented in Section~\ref{sec:model} is to the presence and strength of ionized gas. 
However, the influence of the bar and streaming motions along the arms means this model may not truly reflect the inclination of the underlying stellar or neutral gas disks. 
A deep $g$-band image of HG~190608 obtained with X-shooter on the Very Large Telescope is presented in \cite{Bhandari+20}. 
We carried out a surface photometry analysis on this image using the {\tt isophot} tools within V3.17 of the {\sc stsdas} package in V2.16 of {\sc iraf}\footnote{\url{https://ascl.net/9911.002}}. 
This fits elliptical isophotes to galaxy images, varying the photometric center coordinates ($X,Y$), ellipticity $e$ ($= 1 - b/a$, where $a$ and $b$ are the ellipse semi-major and semi-minor axis lengths, respectively) and position angle $\theta$, using the iterative method of \citet{Jedrzejewski1987}.

A first pass allowed all parameters to vary, but showed the photometric center to be consistent at $<$0.5~pixel.
On the next pass, the center was held fixed, and only $e$ and $\theta$ were allowed to vary. 
Within a radius of $4''$, the bar dominates with large $e$, while beyond $\sim$7$''$ both $e$ and $\theta$ undergo wild swings as the surface brightness drops well below the sky level. 
Between these regimes we find $e = 0.10 \pm 0.02$, corresponding to an inclination $i_{\mathrm{stellar}} = (26 \pm 3)^{\circ}$.

The difference in the derived gas and stellar inclinations is small but significant. A direct correction of the circular rotation speed would imply $v_{\mathrm{circ}} \approx 194 \;\rm km \; s^{-1}$ and $M_{halo}^{\rm dyn} \approx 10^{12.38}$ \msun. However, since the fitting algorithm discussed in the previous section was applied to a gas emission line, a correction using $i_{\mathrm{stellar}}$ may not yield as realistic a solution as separately fitting the stellar kinematics.

\section{FRB Propagation} \label{sec:analysis}

In this section, we analyze our measurements of the local environment to provide context to the FRB measurements.

\begin{table}
\centering
\caption{Relevant and Derived Quantities} \label{tab:derived}
\begin{tabular}{ccc}
\tablewidth{0pt}
\hline
\hline
Parameter & Value & Reference\\
\hline
$\rm{DM_{FRB}}$ & $\dmval$\;\dmunits & \cite{Day+20}\\
$\rm{RM_{FRB}}$ & $\rmvalue \pm \rmerr \, \rmunits$ & \cite{Day+20}\\
$\mmstar$ & $10^{10.4}\;\mmsun$& \cite{Bhandari+20}\\
$v_{\mathrm{circ}}$ & $\vcirc$ & This work\\
$f_{\rm [O\,{\sc III}]}/f_{\rm H\beta}$ & 0.43 & This work\\
$f_{\rm H\alpha}/f_{\rm H\beta}$ & 4.63 & This work\\
$A_V$ & 1.28 & This work \\
$i_{\mathrm{gas}}$ & $37 \pm 3 ^\circ$ & This work\\
$i_{\mathrm{stellar}}$ & $26 \pm 3 ^\circ$ & This work\\
$M_{\mathrm{halo}}^{\rm dyn}$ & $10^{11.96 \pm 0.08}\mmsun$& This work\\
$\rm{DM_{\rm Host}}$ & $149 \pm 45$ \dmunits & This work\\
\sfrsurf & $\mathrm{1.5 \times 10^{-2}\;M_\odot\;yr^{-1}\;kpc^{-2}}$ & This work\\
$\Omega_p$ & $34\pm 6\;\mathrm{km\;s^{-1}\;kpc^{-1}}$ & This work\\
$t_{\mathrm{enc}}$ & $21_{-6}^{+25}$ Myr & This work\\
\hline
\end{tabular}
\end{table}


\subsection{\dmhost} \label{sec:dm}

\cite{Prochaska2019} detail how the dispersion measure of FRBs 
may probe the cosmic web once one accounts for contributions from the 
Milky Way and the FRB host galaxy. 
The latter, in particular, is poorly constrained and may be the
dominant systematic to any such DM analysis \citep{Macquart+20}.
Here we estimate the host contribution, \dmhost, to the total
FRB dispersion measure, \dmfrb, 
from 
  (1) gas in the star-forming ISM of \hgname\  (\dmismhost)
and 
  (2) unseen gas from its galactic halo (\dmhalohost).

For the former, we 
follow the procedure outlined in \cite{Tendulkar2017} 
and references therein \citep{Reynolds1977,Cordes2016}. 
This requires an estimate of the H$\alpha$ emission
measure \emhalpha\ at the FRB position.
From the dust-corrected H$\alpha$ surface brightness
$\sbchalpha$\ 
at the FRB location (see Section~\ref{sec:kcwiflux}), 
we estimate $\memhalpha = \mathrm{293\pm18\;pc\;cm^{-6}}$ \citep{Reynolds1977}. 
We then adopt Equation 5 from \cite{Tendulkar2017} to 
estimate \dmismhost\ in the observer frame:

\begin{equation}
\begin{array}{lcl}
    \mdmismhost & \approx & \mathrm{387\;pc\;cm^{-3}} L_{\mathrm{kpc}}^{1/2} \left[\frac{f_{\mathrm{f}}}{\zeta (1+\epsilon^2)/4}\right]^{1/2}\\
    &&\times\left(\frac{\mathrm{EM}}{\mathrm{600\;pc\;cm^{-6}}}\right)^{1/2} \times (1+z)^{-1}
\end{array}
\end{equation}
where $f_{\rm f}$ is the volume filling factor, $\epsilon$ represents the variation \textit{within} any given cloud of ionized gas due to turbulence, and $\zeta$ is the density variation \textit{between} any two clouds, all over a pathlength of $L_{\rm kpc}$ \citep[]{Reynolds1977, Cordes2016,Tendulkar2017}.

We assume that each ionized cloud along our line of sight has internal density variations dominated by turbulence ($\epsilon$=1) and that there is total variation between clouds ($\zeta$=2). 
Finally, we assume that the FRB resides in the midplane of the spiral arm in which the thin disk scale height is similar to the half width at half maximum (HWHM) of the Milky Way thin disk detected by H\,{\sc i} gas ($L_{\mathrm{kpc}}$ = 0.150; \citealt{KalberlaKerp2009}) and dominates the contribution to DM$_{\mathrm{Host,ISM}}$.
Using our EM$_{\rm H\alpha}$ estimate, we obtain an ISM contribution of $\mathrm{94\pm23}$ \dmunits~with the quoted uncertainty solely from statistical uncertainty in 
the flux measurements.
We estimate a systematic uncertainty of $\sim$ 30 \dmunits~based on our ignorance of $\epsilon,\zeta$, and $L$ and our assumptions to infer $f_{\rm H\alpha}$.  Altogether, we estimate $\mdmismhost = \rm 94 \pm 38$ \dmunits.

This \dmismhost\ estimate has 
adopted the average surface brightness in the $\approx 0.58'' \times 0.87''$
area encompassing the FRB localization.
Figure~\ref{fig:hst} shows that the spiral arms exhibit significant
structure within this region.  Specifically,  we measure a peak
flux in the integration box that is $\approx$1.5 times the average value.
If FRB~190608 occurred at the peak location,
we would estimate a \dmismhost\ value that is $\approx$~20$\%$ higher,
comparable to the statistical error.

\cite{XuHan2015} used the framework of NE2001 to the model dispersion measures of FRBs originating from a spiral galaxy at varying inclinations. The resulting posterior distribution was fit with a skew Gaussian, and for a 40$^\circ$ inclination, they reported a peak DM of $47 \;\mathrm{pc\;cm^{-3}}$ with a right HWHM extending to $90 \;\mathrm{pc\;cm^{-3}}$, slightly beyond the 1-$\sigma$ range of our estimate. 

Ionized gas within the halo of \hgname\ will contribute
an additional factor \dmhalohost\ to \dmhost.
We estimate this contribution as follows.  Starting
from the
stellar mass estimate of \hgname, we implement the abundance
matching technique to infer a halo mass of 
$\mmhalo = 10^{11.9} \, \mmsun$ \citep{Moster+13}. This mass is consistent with the dynamical mass estimated in Section~\ref{sec:model}.
We can then estimate \dmhalohost\ by assuming a density
profile for the halo gas.  For a fiducial estimate,
we assume that the halo has retained all of its cosmic 
fraction of baryons and that $f_{\rm hot} = 75\%$ 
of these are in the halo as
ionized gas. We further assume the modified Navarro-Frenk-White (NFW)
profile described by \cite{Prochaska2019} and
that the halo terminates at a radius of $r = 10$\,kpc, 
with the gas within dominated by the galaxy ISM (and \dmismhost).
Last, we adopt an impact parameter $R_\perp = 6.44$\,kpc
and place the galaxy at the center of the halo such that 
the sightline to FRB~190608 intersects only one half. 

Altogether, we estimate $\mdmhalohost = 49\;\mdmunits$
in our observer frame.
This estimate bears significantly more uncertainty than
the semiempirical \dmismhost\ estimate.
First, errors in the stellar mass and dispersion in 
the abundance matching relation imply an $\approx 0.2$\,dex
uncertainty in \mhalo\ and an $\approx 20\%$ uncertainty in 
\dmhalohost.  Substantially steeper density profiles, strongly
disfavored by simulations of galaxy formation and 
low X-ray emission from spiral galaxies, would allow 
for a \dmhalohost\ up to 100\%\ larger. 
Lastly, the halo may be deficient in baryons with 
$f_{\rm hot} \ll 0.75$.  Altogether, we suggest
a range $\mdmhalohost = 30 - 80\;\mdmunits$ ($55 \pm 25 \; \mdmunits)$.

Combining with our empirical estimate for \dmismhost,
we estimate $\mdmhost = 149 \pm 45\;\mdmunits$.
This is $\approx 30-60\%$ of \dmfrb\ and likely 
at least 50\%\ of the value corrected for the Galaxy.

\subsection{Scattering} \label{sec:turb}

\citet{Day+20} measured the scattering timescale of the burst by fitting a Gaussian intrinsic profile convolved with the approximately exponential tail expected of temporal smearing by an inhomogeneous cold plasma.  The burst profile was fit as a function of frequency across seven subbands covering the bandwidth range 1105-1433 MHz. The authors found that the observed frequency-dependent burst profile can be adequately fit by a single Gaussian component of constant width modified by frequency-dependent scattering, where the scattering timescale follows a power law  with the frequency as \dt = $(\dtval \pm \dterr) \, (\nu/1280\,{\rm MHz})^{-3.5 \pm 0.9}$ ms \citep{Day+20}.  The index of the scattering and its error are consistent with scattering by Kolmogorov turbulence at the 1$\sigma$ level (index = -4.4) or with scattering where the diffractive scale is below the inner (dissipation) scale of the turbulence at the 1$\sigma$ level (index = -4.0). The index is consistent to 2$\sigma$ with the average index measured for pulsar lines of sight in our Galaxy \citep{Bhat2004}. We interpret the result in terms of scattering by a Kolmogorov spectrum of density inhomogeneities for ease of comparison against the properties of the interstellar medium of our own Galaxy.

The measured scattering timescale cannot be explained by scattering in our Galaxy, being 3-4 orders of magnitude larger than expected at the observed Galactic latitude of $-48.6^\circ$ \citep{Cordes}.

However, the magnitude of the scattering is problematically large if attributed to the ISM of the host galaxy.  One would expect the host galaxy to contribute an amount of temporal smearing comparable to the Milky Way given the low inclination of \hgname~ ($\iangle$), the burst location in an outer spiral arm, and the fact that the gas mass of the host is comparable to the Milky Way.  If the burst is in the midplane of the host galaxy, its sightline would be equivalent to sightlines in the Milky Way at Galactic latitudes $\gtrsim 50^\circ$, for which the scattering measure is ${\rm SM} < 10^{-3.7}$\,kpc\,m$^{-20/3}$.

Nonetheless, the derived scattering measure is ${\rm SM} = 1.4  (D_{\rm eff}/ 1$\,${\rm kpc})^{-5/6}\,$kpc$\,$m$^{-20/3}$, where we have conservatively assumed a fiducial effective distance to the scattering medium of $D_{\rm eff} \sim 1\,$kpc from the burst location. 

The relationship between dispersion measure and scattering measured for pulsars in the Milky Way \citep{Bhat2004,Krishnakumar2015} provides an independent means to estimate the expected amount of scattering based on the estimated host DM contribution.  Using the estimated upper bound of $\mdmismhost = \mathrm{94\pm38\;pc\;cm^{-3}}$ and ignoring any negligible contribution from the much sparser gas in the halo, we estimate a temporal smearing timescale of $3~\mu \,$s.

The discrepancy between the observed smearing timescale and these various estimates of the expected host galaxy ISM contribution leads us to conclude either that:
\begin{itemize}
        \item the scattering instead arises in an exceptionally turbulent and dense medium associated either with the burst/circumburst medium or a dense H\,{\sc ii} region in the spiral arm of the host galaxy or;
        \item the scattering arises at cosmological distances due to some turbulent intervening structure, where the large effective scattering distance alleviates the requirement of a large scattering measure (since $\tau \propto D_{\rm L} D_{\rm LS}/D_{\rm S}$, where $D_{\rm S}$ and $D_{\rm L}$ are the angular diameter distances to the source and scattering plane, respectively, and $D_{\rm LS}$ is the distance between the source and scattering plane).
\end{itemize}
We discuss each of these possibilities in turn.

(i) We consider whether the burst could originate in an especially turbulent and dense region.  Any such association plausibly confines any putative scattering region to $D_{\rm eff} < 10\,$pc, which requires ${\rm SM} > 65\,$kpc\,m$^{-20/3}$.  Scattering that is yet more local to the burst would imply an even higher constraint on the scattering measure. We remark that the observed scattering is 2.5 orders of magnitude greater than that observed in any analogous system observed in our Galaxy: the most extreme scattering environment observed in the Milky Way is associated with an energetic neutron star -- the Crab nebula -- whose scattering reaches values of $600\,\mu$s at 610 GHz \citep{McKee+2018}, equivalent to 15 $\mu$s at 1.4\,GHz.

The scattering could instead originate from the chance intersection of a dense, turbulent H\,{\sc ii} region associated with the line of sight. We regard this as the most plausible option. However, we note that the scattering measure is comparable to the highest values encountered in our own Galaxy -- notably those toward the Galactic Centre, whose scattering is attributed to an H\,{\sc ii} region $\approx 2\,$kpc from Earth \citep{Dexteretal2017}.

(ii) The scattering could be associated with structure in the cosmic web along the line of sight. 
Regarding the diffuse intergalactic medium, \cite{Simha+20} have performed a reconstruction
of the cosmic web and report 
no evidence for a large concentration of matter along this line-of-sight.
Regarding scattering from the gas in intervening galactic halos,
\cite{Simha+20} reported that only a single halo is intercepted along the
sightline (at $z=0.09$) and estimated its scattering contribution to 
be $\tau_{\rm halo} \lesssim 0.1$\,ms.  Therefore,
it is improbable that the cosmic web dominates the inferred scattering
of FRB~190608.

\subsection{Rotation Measure} \label{sec:rm}

FRB~190608 has one of the larger rotation measures
recorded for an FRB to date \citep[RM~$= \rmvalue \pm \rmerr \, \rmunits$,][]{Day+20},
requiring a highly magnetized plasma
along the sightline. The estimate for 
the Milky Way Galactic halo
at this high Galactic latitude
is ${\rm RM}_{\rm MW} = -25 \pm 8 \, \rmunits$
\citep{Oppermann2012};  therefore, we expect the signal to be
dominated by an extragalactic component.

We identify four possible origins for the high observed RM:

\begin{itemize}
    \item a foreground galaxy or halo along the line of sight, 
    \item the host galaxy ISM, without a large-scale magnetic field,
    \item large-scale magnetic fields in the host disk and/or halo, or
    \item a dense and turbulent medium in which the burst could potentially be produced.
\end{itemize}{}

We examine each scenario below.

(i) Highly magnetized foreground material could plausibly produce the observed high RM. However, \citet{Simha+20} estimated a negligible contribution to the RM ($<1\;\rmunits$) from their analysis of the intervening halos along the FRB~190608 sightline, leaving the host and local environment as the most likely suspects.

(ii) A high degree of regularity, over large scales, in the magnetic fields threading the host galaxy ISM could be responsible for the high observed RM. Analysis of magnetic field strengths (via rotation measures) from spiral arms in the outer disk of the Milky Way yield a characteristic $|{\rm RM}| \approx 180 \, \rmunits$ \citep{Brown2001}. However, we note this estimate was based on observations made at low galactic latitudes, and this need not apply to the observed rotation measure in a roughly face-on spiral galaxy as in the case of \hgname. The findings in \citet{Fletcher2011} of the magnetic field in M51, a face-on spiral galaxy
\citep[$i = (22 \pm 5)^\circ$;][]{Colombo2014}, are more relevant. The authors found rotation measures outside of the galaxy center in the disk to be generally $|\rm RM|<100\;\rmunits$ and the total magnetic field strengths to be strongest in the interarm regions \citep[10-15$\rm~\mu G$;][]{Beck2015} rather than in the arms. We note that the authors found no organizing field or pattern to the RM, i.e. no large-scale magnetic field present.

M51 also has an $\Sigma_{\rm SFR}$ of 1 order of magnitude higher than \hgname~\citep{Leroy2017}. This corresponds to a lower expected thermal electron density, lower magnetic field, and a lower RM in \hgname. Furthermore, turbulent cells in the host ISM along the line of sight would reduce the contributed RM by a factor of $\sqrt{N}$, where \textit{N} is the number of cells. For cell sizes of $\sim 50$ pc \citep{Fletcher2011}, we can expect the RM contribution of the ISM to be reduced by a factor of $\sim 2$.

Thus, for the host galaxy ISM to dominate the observed RM, the magnetic field would have to be somewhat larger (and/or remain ordered on larger physical scales) than that of the Milky Way or M51, albeit only by a factor of a few.

(iii) While M51 lacks a large-scale magnetic field, \citet{Mora-Partiarroyo+19} found that a phenomenon known as ``magnetic ropes" could produce such a field, as in the edge-on spiral galaxy NGC 4631. A regular magnetic field with a strength of $\approx$ 4 $\mu$G oriented out of the plane of the galaxy was observed, with $|\rm RM|$ as high as 400 $\rmunits$ in some regions. If a similar phenomenon is present in \hgname, we can estimate the magnetic field strength parallel to the sightline as:

\begin{equation}
\begin{array}{lcl}
  B_\parallel & = & 9.2\,{\rm \mu G}
      \; \ltp \frac{\rm RM}{\rmvalue \, \rm \rmunits}\rtp{}
      \; \ltp \frac{n_e}{0.05 \, \rm cm^{-3}}\rtp^{-1} \\
      && \times \; \ltp \frac{L}{1000 \, \rm pc}\rtp^{-1}
\end{array}
\end{equation}
Here, we have adopted a fiducial electron density for the ISM and assumed a path length through both the thin and thick disk. This magnetic field strength is stronger than that observed in NGC 4631 but of the same order. However, we cannot infer that such a phenomenon is present without a study similar to \cite{Fletcher2011} or \cite{Mora-Partiarroyo+19} conducted on \hgname.

(iv) Finally, we consider that a dense and turbulent medium associated with the FRB source (i.e. a circumburst medium) gives rise to the high RMs. For a region with $n_e$ = 5 cm$^{-3}$ and $L$ = 10 pc, we could expect the same parallel field strength as the scenario above. Though this would be an unusually powerful magnetic field, it is not unprecedented, especially in the field of FRBs.

In the case of the FRB 121102, the first FRB detected to repeat, \citet{Michilli2018} reported a decaying source frame rotation measure of $\rm RM_{\rm src} \sim 10^5\;\rmunits$ and a magnetic field parallel to the line of sight on the order of $\sim$mG, much stronger than what we have considered here for FRB 190608. \citet{MaraglitMetzger2018} proposed that the progenitor of FRB 121102 could be a young magnetar embedded in a magnetized nebula. While the observables for FRB 190608 do not indicate as unique an environment, given the unknown nature of FRBs, we consider it possible that a magnetic environment related to the burst or circumburst medium gives rise to the high rotation measure.

We conclude that the most likely explanation for the observed rotation measure is a combination of contributions from the host galaxy ISM and the burst/circumburst medium.

\section{Discussion} \label{sec:discussion}

\subsection{The Local Environment of FRB~190608} \label{sec:locenv}

We now consider the local environment of FRB~190608 and compare
its properties with the overall properties of the galaxy.
First, the FRB localization is coincident with one of the
two prominent spiral arms of the galaxy.
Defining these arms by the luminous UV emission (i.e.\ the contours
in Figure~\ref{fig:hst}),
we estimate a chance coincidence of 20\%\ for
an event occurring within 3 effective radii of
the galaxy center. Therefore, a chance association is unlikely.


\begin{figure}
\centering
\includegraphics[width=0.49\textwidth]{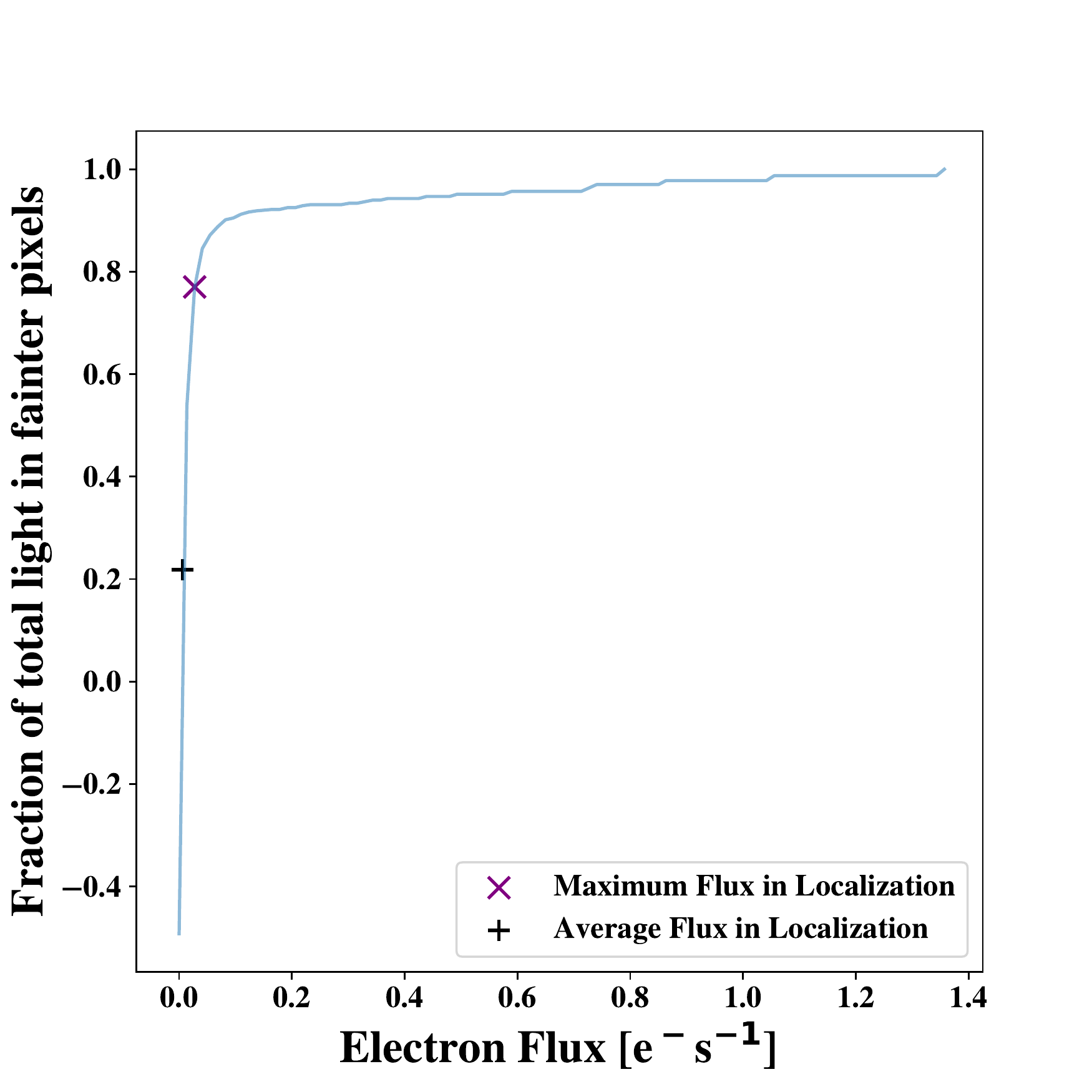}
\caption{The figure shows the fraction of total light in pixels fainter than a given electron flux value - with fraction of light as a function of electron flux. This can be used to show the relative brightness of the FRB localization. We find that the maximum brightness in the area of localization is one of the more UV-bright pixels in the image but not the brightest. Here we show the fraction of light in pixels fainter than the maximum flux is approximately 0.78 -- i.e, $\approx78\%$ of the light in the image is in pixels fainter than the localization maximum, indicated by the blue cross. The burst may have occurred in a relatively bright star-forming region but not the most luminous as found with GRB local environments.}
\label{fig:lightcomp}
\end{figure}


While an association of FRB~190608 to a star-forming region
within a spiral arm of HG~190608 is probable,
we emphasize that the event did not occur in the most active
star-forming region.
On the other hand,  Figure~\ref{fig:lightcomp} shows
a histogram of the cumulative pixel fluxes from the
HST/UVIS image. Using a method similar to that used for GRB analysis \citep{Fruchter06} to compare GRB environments to supernovae environments, we calculated the fraction of total light in pixels fainter than the average and maximum pixel brightness at the FRB localization. 

These fractions are calculated using 

\begin{equation}
    f =\frac{\Sigma(F_{i} < limit)}{\Sigma F_{i}}
\end{equation}
\noindent
where $F_{i}$ is a pixel flux, and the limit is some number between 0 and the maximum flux in the entire image. This can sometimes result in a negative fraction as seen in Figure~\ref{fig:lightcomp} because of negative flux values dominating the sum in the numerator. We plotted markers at specific limits associated with the FRB localization.

We estimate that, for the average flux 
of $\rm{0.00638\; e^-\;s^{-1}}$ in an elliptical aperture with axes equal to $\sigma_{RA}$ and $\sigma_{Dec}$, the fraction of light in pixels fainter than this is 0.22. 
For the maximum flux in the localization ellipse 
($\rm{0.0276\;e^-\;s^{-1}}$), 
the fraction of light in pixels fainter than this value is 0.780. (Figure \ref{fig:lightcomp}).

Thus, the event did not occur in the brightest star-forming region of the galaxy. This contrasts with long-duration GRBs whose progenitors
track the most luminous UV emission of their host galaxies
\citep[e.g.][]{Fruchter06,Lyman2017}.

We can estimate the SFR surface density ($\Sigma_{\mathrm{SFR}}$) at the FRB position from \muchalpha. Using the canonical $L_\mathrm{{H\alpha}}$-SFR relationship from \cite{Kennicutt1998}, 
from Section \ref{sec:kcwiflux}, 
we estimate $\Sigma_{\mathrm{SFR,FRB}} = 1.5 \times \; \mathrm{10^{-2}\;M_\odot\;yr^{-1}\;kpc^{-2}}$. 
This value is similar to the average SFR estimated from the $3''$ SDSS 
fiber covering the inner regions of the galaxy.
Similar to the UV flux analysis in Figure~\ref{fig:lightcomp},
we infer no enhanced star-formation at the FRB location relative to
other areas in the inner few kpc.
However, we note that there is no favored environment for a progenitor when compared with overall galaxy properties \citep{Bhandari+20}.

We have searched the data for signatures of disturbance
or anomalous emission in the environment of FRB~190608.
We derive a modest \sfrsurf\ consistent with the emission
along the spiral arms of the galaxy.
Similarly, we derive an H$\beta$/[O\,{\sc iii}] ratio at the FRB,
which is comparable to the remainder of the galaxy 
(excluding the AGN-dominated nucleus).
Finally, the gas kinematics closely track the overall
rotation of the galactic disk, and there is no excess 
velocity dispersion.

\subsection{Lindblad Resonance Analysis} \label{sec:pattern}
The origin and mechanism of galactic spiral structure frequently observed in the present-day universe remains an open question. The theory of kinematic density waves driven by resonances first introduced in \citet{Lindblad1959} suggests that an initial radial perturbation in the disk gives rise to spiral structure that is well described by a wave moving at a defined pattern speed, $\Omega_p$. According to Lindblad, the epicyclic motions of stars, denoted as a frequency $\kappa$, in response to the radial perturbation and the rotation frequency of the disk, $\Omega\; = v_{rot}/R$, generate global resonances now called Lindblad resonances. By identifying the radii at which these resonances occur, it is possible to infer the global pattern speed of the spiral density wave \citep[see][for more detail]{LBK1972,Kalnajs1973,Wielen1974,BinneyTremaine}.

Kinematic density wave theory is only one of many theories to explain spiral structure. \citet{LinShu} developed the theory of quasi-stationary density waves set up by a gravitational instability in the galactic disk, and take the gravity of the spiral arms into consideration. Here, the stability of the wave is not limited to resonant frequencies, and hence does not depend on global resonances as in the case of kinematic density wave theory. Spiral arms, particularly for grand design galaxies \citep{EE1982} are also thought to be generated, or at least enhanced, through tidal interactions with companion galaxies \citep[see][and references therein]{TT1972,Kendall2011,Oh2015}. However this may not explain the spiral structure of HG 190608 which lacks a clear nearby companion \citep{Simha+20}. The last theory we mention is from \citet{SellwoodCarlberg} who propose that a few nonlinear modes generated by self-excited instabilities in the disk are the origin of spiral structure. Here, each instability gives rise to temporary spiral behavior, but then scattering at Lindblad resonances will eventually generate future instabilities and thus together can account for long-lived spiral structure. As \citet{DobbsBaba2014} note, these various mechanisms need not necessarily compete, and may play a role together for a given galaxy.

Spiral arms consistently display enhanced star formation in optical light emitted by young, massive stars and in the ultraviolet as in Figure \ref{fig:hst}. As the rotating wave characterized by the pattern speed overtakes the slower, out-of-spiral material in the galactic disk, the gas is shocked and compressed as it moves into and through the spiral arm. This compression is what eventually leads to enhanced star formation. Assuming that (linear) kinematic density wave theory applies to HG 190608, we can infer the bar-driven pattern speed by the identified second harmonic (4:1) Lindblad resonance at the star-forming ring at $\sim$3 kpc. The bar-driven pattern speed and the spiral density wave speed may generally be different. However, given that the FRB position is near the onset of the spiral arm, which emanates directly from the ends of the bar, the two pattern speeds ought to be similar at this radius lest the arms become ``disconnected" from the bar. We aim to use this analysis to determine the maximum time since the outermost radial position of the FRB localization could have encountered the leading edge of the spiral density wave. This will allow us to place an upper limit on the age of a stellar-type progenitor for an FRB assuming that spiral arm compression induced the formation of the progenitor.

Returning to Lindblad's resonance theory, various resonances correspond to different $\mathrm{\Omega\;\pm\;\kappa/n}$ curves on a plot of $\Omega$ vs. $R$, where $n$ is an integer value corresponding to the resonance and $R$ is the galactocentric distance. For example, the inner Lindblad resonance (ILR) corresponds to $\mathrm{\Omega\;-\;\kappa/2}$ (2:1), the outer Lindblad resonance (OLR) corresponds to $\mathrm{\Omega\;+\;\kappa/2}$ (2:1), and the corotation resonance at which both the disk and bar rotate with the same velocity, namely $\Omega_p$. The 4:1 inner second harmonic resonance marked by the inner ring encircling the bar corresponds to $\mathrm{\Omega\;-\;\kappa/4}$. Although only hinted at in the UV
(Figure~\ref{fig:hst}), the $g$-band image of HG~190608 in \citet{Bhandari+20}
appears to show the outer spiral arms wrapping around into an outer pseudoring
that would correspond to the OLR \citep[e.g.][]{Buta1991}.
The pattern speed of the bar should therefore be the frequency at which each resonance's galactocentric distance intersects their respective frequency curve. Further, we expect that the frequency of intersection be the same for all resonances, hence reflecting a single pattern speed.

We begin by computing the epicyclic frequency, $\kappa$, by assuming a small radial perturbation with linear solutions in the radial and azimuthal directions (resulting in a solution for simple harmonic motion):
\begin{equation}
    \kappa = \sqrt{2}\;\Omega\ltp1+\frac{1}{2}\ltp\frac{R}{\Omega}\;\frac{d\Omega}{dR}\rtp\rtp
\end{equation}
Here, $\frac{d\Omega}{dR}$ is calculated with \texttt{numpy.gradient}, an algorithm that uses central differences for interior points and one-sided differences at the boundaries \citep{numpy}.

\begin{figure}
\centering
   \includegraphics[width=0.47\textwidth]{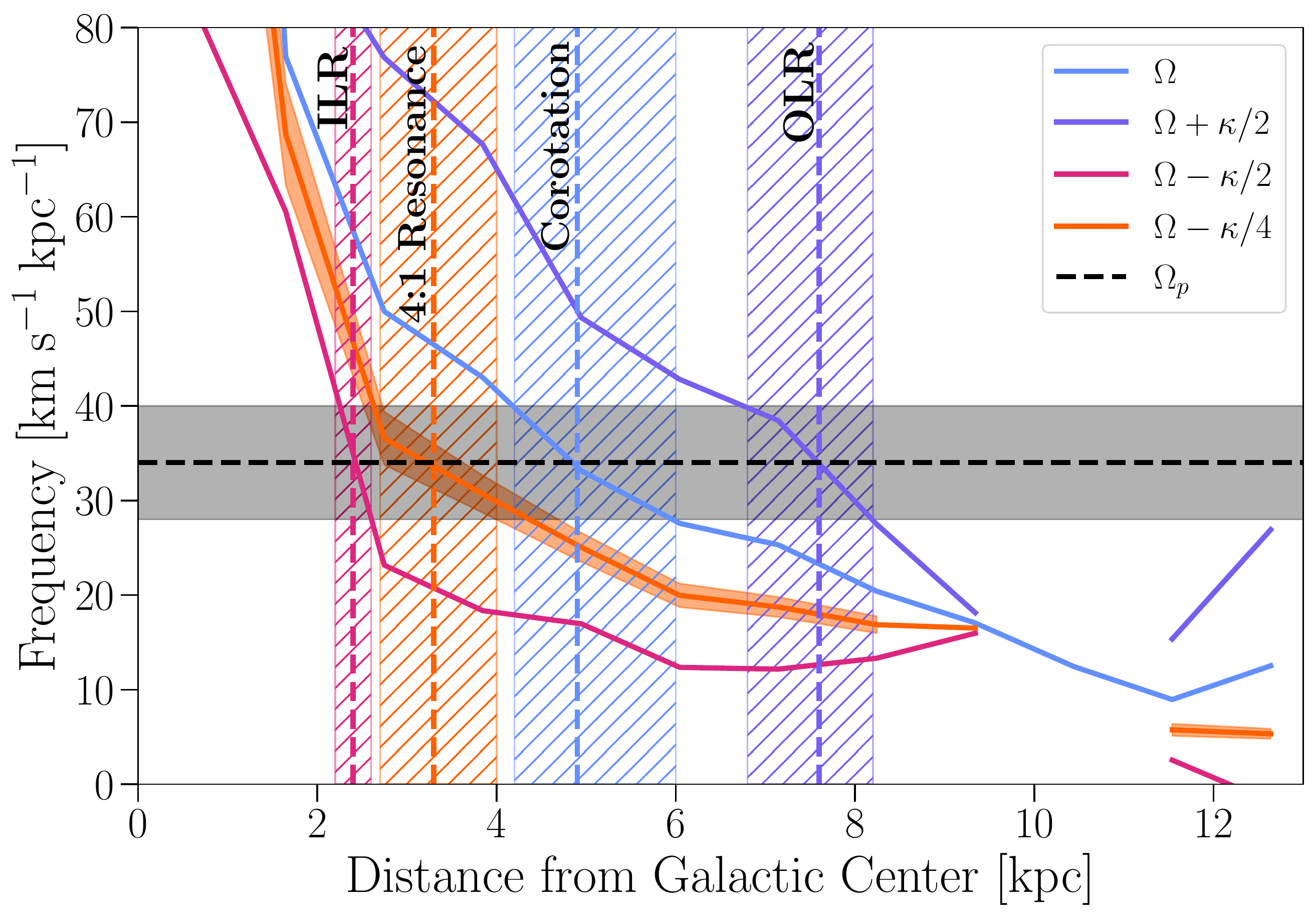}
\caption{Rotation frequency curves produced using the rotation curve generated by \texttt{$^{3{\rm D}}$Barolo}. The 4:1 resonance corresponds to the orange $\mathrm{\Omega\;-\;\kappa/4}$ curve, the OLR to the purple $\mathrm{\Omega\;+\;\kappa/2}$ curve, the ILR to the magenta $\mathrm{\Omega\;-\;\kappa/2}$ curve, and corotation to the blue $\Omega$ curve. The orange hatched region is the 1.3 kpc width of the 4:1 resonance ring, and the orange shaded region is the uncertainty in the associated frequency curve assuming a 10 $\mathrm{km\;s^{-1}}$ uncertainty in the rotation curve. The black horizontal dashed line and shaded region denote the pattern speed of $\mathrm{\Omega_p\;=\;34\pm 6\;km\;s^{-1}\;kpc^{-1}}$. The other hatched regions indicate the positional uncertainties of the remaining resonances based on the inferred pattern speed.}
\label{fig:omega}
\end{figure}

Next, we plot the frequency curves for the ILR, the 4:1 resonance, corotation, and the OLR as a function of the galactocentric distance in Figure \ref{fig:omega}. The vertical lines represent the inferred radial positions of these resonances. As shown in the figure, the relevant curves intersect their respective resonances at a pattern speed of $\Omega_p= 34\pm6\;\mathrm{km\;s^{-1}\;kpc^{-1}}$. The uncertainty on $\Omega_p$ is determined by assuming a 10 km s$^{-1}$ uncertainty in the rotation curve as well as considering the finite width of the ring, $\approx$1.3 kpc. From the pattern speed, we can infer that the ILR might occur at $2.4\pm0.2$ kpc, which we note is just outside of the central-most bright region in the HST/UVIS image but otherwise does not have an observational signature. Corotation would then occur in the regime of $4.9_{-0.7}^{+1.1}$ kpc, where the ring ends and the spiral arms begins, consistent with expectations for a barred galaxy. Lastly, the OLR is at $7.6_{-0.8}^{+0.6}$ kpc just beyond the extent of the inner spiral arms, but these open out to form more of a pseudo-ring that could be associated with this resonance \citep{Ryder1996}. Our surface photometry analysis of the X-shooter image in Section \ref{sec:ospa} indicates that the bar radius is within 4$\arcsec$ (8.8 kpc), and could be as low as 3$\arcsec$ (6.6 kpc) when considered along with the apparent bar in the HST image. Theory and observations suggest that the ratio of the corotation radius to the bar radius be $R_{\rm CR}/R_{\rm bar}\geq1$ \citep{Contopoulos1980, Debattista2000, Aguerri1998}, while our results are in the range of $R_{CR}/R_{\rm bar} = 0.48-0.91$. Ratios near (or below) unity would suggest a very fast bar. While this is atypical, it does not on its own disprove that resonance rings are present in \hgname.

At the projected distance of the FRB ($\approx$ 6.4 kpc) our results imply that the bar-induced wave moves at a speed of $219_{-39}^{+32}$ $\mathrm{km\;s^{-1}}$. From the rotation curve, the inclination-corrected rotation velocity is 167 $\mathrm{km\;s^{-1}}$ implying the density wave moves $\approx$52 $\mathrm{km\;s^{-1}}$ faster than the stars and gas at the same radial position. Lastly, we infer a maximum distance from the wave front to be the diameter of the uncertainty in the localization, which is coincidentally the approximate width of the spiral arm as seen in the UV, about 1.1 kpc in physical distance. We calculate that the maximum time since the FRB position could have encountered the wave front to be $t_{\mathrm{enc}} = 21_{-6}^{+25}$ Myr.


If (a) FRB 190608 indeed occurred at the furthest distance in the localization from the wave front; (b) we assume our application of linear resonance theory is plausible and (c) consider a stellar-type progenitor formed by gas compression at the wave front, then an age of 21 Myr would make a young magnetar scenario possible. A 10 $\mmsun$ star has a typical lifetime of $\sim$30 Myr with higher mass stars having shorter lifetimes still, and young (active) magnetars are thought to live for only $\sim$10$^4$ years \citep{Colpi2000,Beniamini2019}. An age of 21 Myr or the lower limit of 15 Myr are possible for a magnetar, but the upper limit of 46 Myr would be unlikely as it is beyond the expected age range for such an object.

However, \citet{Aramyan2016} showed from their study of supernovae offsets in spiral galaxies that the peaks of spiral arms are the most likely sites of star formation induced by the density wave. If this holds for FRB 190608, then we might expect that the stellar-type progenitor to be born at roughly half $t_{\mathrm{enc}}$, at approximately 10 Myr (shorter still at the lower limit) and thus we cannot exclude the young magnetar hypothesis. We also note that magnetars can be kicked during their supernovae explosions and have been observed to have velocities of hundreds of $\mathrm{km\;s^{-1}}$ while numerical simulations suggest even higher velocities of $\sim$1000 $\mathrm{km\;s^{-1}}$ are possible \citep{Sawai2008,Deller2012}. However, even if a magnetar were to be kicked in the direction of galactic rotation, we calculate a maximum offset of about 10 pc over its active lifetime in the spiral frame and thus this effect is subdominant to the results discussed here.

The \texttt{$^{3{\rm D}}$Barolo} rotation curve does not have a formal uncertainty, but if our constant velocity fit from \texttt{qubefit} is valid at the 4:1 resonance and the FRB position, then those results would also suggest a pattern speed of $\mathrm{\Omega_p\;=\;34\pm 8\;km\;s^{-1}\;kpc^{-1}}$ and thus a similar $t_{\mathrm{enc}}$ as the estimate above. This agreement is due to the fact that the two rotation curves are coincident at the position of the 4:1 resonance (see Figure \ref{fig:vrot}). Given the positional uncertainty and the low resolution of our rotation curve, we offer this analysis as a proof of concept for future studies of FRBs in spiral arms when resonance rings can be identified.

\subsection{Comparison to the host of FRB 180916.J0158+65} \label{sec:compare}

The detection of \hgname~offers the opportunity to compare galaxy properties with the spiral host galaxy of FRB 180916.J0158+65, a repeating FRB whose localization was reported in \cite{Marcote2020}.

FRB 180916.J0158+65 was associated with a star-forming clump in a $z$=0.0337 galaxy with the redshift identified from a Gemini-North long-slit spectrum. Emission lines from [N\,{\sc ii}], \ha, and [O\,{\sc iii}] available in their spectrum suggest the host is most likely a star-forming galaxy when considered on a BPT diagnostic plot \citep{BPT81}.
This is in contrast to \hgname, which \cite{Bhandari+20} identify as a LINER galaxy from the \verb|pPXF|-processed SDSS spectrum.

The V-shaped star-forming clump associated with FRB 180916.J0158+65 suggests a perturbed environment with a projected size of about 1.5 kpc. 
\cite{Marcote2020} suggest that this region is likely the result of an interaction with a satellite dwarf galaxy or between multiple star-forming regions. \hgname~has no remarkable features at the FRB position indicative of a history of galactic interactions. 
\cite{Marcote2020} also estimate a star formation surface density on the order of $\mathrm{10^{-2}\;M_{\odot}\;yr^{-1}\;kpc^{-2}}$, similar to our estimate
at the location of FRB~190608.

In a host galaxy study of FRB 180916.J0158+65 with IFU data from the MEGARA spectrograph and HST imaging, \citep{Tendulkar2020} completed a similar analysis as in Section \ref{sec:pattern} and found that the FRB was offset from the nearest identified star-forming region by about 250 pc. The authors conclude that even with a high kick velocity, a magnetar progenitor would have taken 0.25 Myr to travel to the observed position. Here, our position and velocity uncertainties only allow us to place an upper limit for the stellar-type progenitor to be 40 Myr, but both results may challenge the magnetar hypothesis for these FRBs.

FRB 190608 also has a Faraday rotation measure 3$\times$ higher than FRB 180916.J0158+65 \citep[RM $= -114.6 \pm 0.6\; \rmunits$,][]{CHIME+19}. While the RM of 180916.J0158+65 is more consistent with observations of M51 \citep{Fletcher2011}, we cannot exclude the possibility that both are dominated by contributions from the host ISMs.

Despite the fact that both FRB hosts are spiral galaxies, they have distinct galactic and FRB properties. The same can be said when considering the low-metallicity dwarf galaxy hosting the repeating FRB 121102 \citep{Tendulkar2017}. From these global properties, we can conclude that FRBs, whether repeating or not, can occur in very different galactic environments. This may support the burgeoning hypothesis that FRBs have different origins and that there are distinct populations of bursts. Our results alone, however, cannot conclusively determine to which population FRB 190608 belongs.

\section{Concluding Remarks} \label{sec:conclusion}

We have presented an analysis of the spiral galaxy hosting FRB~190608 in order to study the local environment with a focus on observed propagation effects of the burst. We summarize our primary results as follows:

\begin{itemize}
    \item While the FRB is coincident with a bright star-forming region of the galaxy identified in the UV, it is not the \textit{brightest} region, in contrast
    to the majority of long-duration GRB environments.
    
    \item From the inferred H$\alpha$ flux, we estimate the host galaxy ISM dispersion measure contribution to be $\mdmismhost = \mathrm{94\pm38\;pc\;cm^{-3}}$. From the stellar and dynamical mass measurements, we estimate a halo contribution of $\mdmhalohost = 55 \pm 25$ \dmunits, for a total \dmhost$ = 145 \pm 45$ \dmunits.
    
    \item The large observed scattering timescale of the burst is most likely due to a dense, turbulent H\,{\sc ii} region within the galaxy that is intersecting our sightline. An exceptional environment with no Milky Way analog would be necessary if the scattering occurred locally, or very local to the burst.
    
    \item A Faraday rotation measure of $\rmvalue \pm \rmerr \; \rmunits$ would be unusually high to attribute solely to the host ISM, but not implausible. The local environment of the FRB would need to be highly magnetized and/or dense (compared to the ISM) if it were the source of the high RM. We consider the most likely case is that both the local environment and ISM contribute to the observed RM.
        
    \item From the rotation curve produced by \texttt{$^{3{\rm D}}$Barolo} and the radial position of the 4:1 resonance star-forming ring identified in the UV, we estimate a bar-induced pattern speed of $\Omega_p= 34\pm 6\;\mathrm{km\;s^{-1}\;kpc^{-1}}$ assuming linear resonance theory applies. We use the maximum distance from the leading edge of the spiral arm within the positional uncertainty of the FRB to estimate a maximum age of a stellar-type progenitor born from compression-induced star formation of $t_{\mathrm{enc}} = 21_{-6}^{+25}$ Myr.
    
    \item \hgname~is similar to the host of FRB \newline 180916.J0158+65 with comparable $\Sigma_{\rm SFR}$ and morphology. However, we can identify no morphological, kinematic,
    nor emission perturbations at the location of FRB~190608.
\end{itemize}{}

In the absence of optical or other higher-energy counterparts to detected FRBs, galactic host analysis remains one of the most informative paths forward to identifying progenitors. We will continue our multiwavelength investigation of \hgname\ with \hst/IR results in future work on behalf of the \textit{Fast and Fortunate for FRB Follow-up} ($\rm{F^4}$)\footnote{\url{ucolick.org/f-4}} collaboration
along with similar analysis for galaxies hosting well-localized FRBs.

\section*{Acknowledgements}
Based on observations collected at the European Southern Observatory under ESO programmes 0102.A-0450(A) and 0103.A-0101(B).
Authors J.S.C., R.A.J., S.S., A.M., J.X.P.,
N.T., and K.H., as members of the Fast and Fortunate for FRB
Follow-up team, acknowledge support from 
NSF grants AST-1911140 and AST-1910471.
This work is supported by the Nantucket Maria Mitchell Association. R.A.J. and J.S.C. gratefully acknowledge the support of the Theodore Dunham, Jr. Grant of the Fund for Astrophysical Research. J.S.C. would like to thank Rainer Beck and Debra M. Elmegreen for their input on the magnetic fields and Lindblad resonances in spiral galaxies, respectively, as well as the reviewer for their helpful comments which greatly improved this paper.
K.W.B., J.P.M, and R.M.S. acknowledge Australian Research Council (ARC) grant DP180100857.
A.T.D. is the recipient of an ARC Future Fellowship (FT150100415).
L.M. acknowledges the receipt of an MQ-MRES scholarship from
Macquarie University.
R.M.S. is the recipient of an ARC Future Fellowship (FT190100155)
N.T. acknowledges support by FONDECYT grant 11191217. K.E.H. acknowledges the support by a Project Grant (162948–051) from The Icelandic Research Fund.

The Australian Square Kilometre Array Pathfinder is part of the Australia Telescope National Facility, which is managed by Commonwealth Scientific and Industrial Research Organisation (CSIRO). 
Operation of Australian Square Kilometre Array Pathfinder (ASKAP) is funded by the Australian Government with support from the National Collaborative Research Infrastructure Strategy. ASKAP uses the resources of the Pawsey Supercomputing Centre. Establishment of ASKAP, the Murchison Radio-astronomy Observatory and the Pawsey Supercomputing Centre are initiatives of the Australian Government, with support from the Government of Western Australia and the Science and Industry Endowment Fund. 
We acknowledge the Wajarri Yamatji as the traditional owners of the Murchison Radio-astronomy Observatory site. 

Spectra were obtained at the W. M. Keck Observatory, which is operated as a scientific partnership among Caltech, the University of California, and NASA. The Keck Observatory was made possible by the generous financial support of the W. M. Keck Foundation. The authors recognize and acknowledge the very significant cultural role and reverence that the summit of Maunakea has always had within the indigenous Hawaiian community. We are most fortunate to have the opportunity to conduct observations from this mountain. 

The NUV data are based on observations with the NASA/ESA Hubble Space Telescope obtained [from the Data Archive] at the Space Telescope Science Institute, which is operated by the Association of Universities for Research in Astronomy, Incorporated, under NASA contract NAS5- 26555. Support for Program number 15878 was provided through a grant from the STScI under NASA contract NAS5- 26555.

\software{scipy \citep{scipy},  
          numpy \citep{numpy}, 
          matplotlib \citep{matplotlib},
          qubefit \citep{Neeleman2019},
          $^{3{\rm D}}$Barolo \citep{DiTeodoro2015},
          CWITools \citep{CWITools},
          KCWI\;Data\;Reduction\;Pipeline \citep{Morrissey2018},
          photutils \citep{photutils},
          astropy \citep{astropy}}
          
\bibliography{references.bib}

\end{document}